\documentclass[sigconf,10pt]{acmart}

\usepackage{graphicx}
\usepackage{balance}  
\usepackage{comment}
\usepackage{accents}
\usepackage{lambda,cc}
\usepackage{mathtools} 
\usepackage{array,multirow}
\usepackage{arydshln}
\usepackage{xcolor}
\usepackage{mathpartir}
\usepackage{hyperref}
\usepackage[normalem]{ulem}
\usepackage{tikz}
\usepackage{tikz-cd}
\usetikzlibrary{decorations.pathmorphing}
\usepackage{wrapfig}

\usepackage{listings} 

\newcommand{\tabref}[1]{\hyperref[tab:#1]{Table~\ref*{tab:#1}}}
\newcommand{\figref}[1]{\hyperref[fig:#1]{Figure~\ref*{fig:#1}}}
\newcommand{\secref}[1]{\hyperref[sec:#1]{\S~\ref*{sec:#1}}}
\newcommand{\defref}[1]{\hyperref[def:#1]{Definition~\ref*{def:#1}}}
\newcommand{\appref}[1]{\hyperref[app:#1]{Appendix~\ref*{app:#1}}}
\newcommand{\chref}[1]{\hyperref[ch:#1]{Chapter~\ref*{ch:#1}}}
\newcommand{\thmref}[1]{\hyperref[thm:#1]{Theorem~\ref*{thm:#1}}}
\newcommand{\lemref}[1]{\hyperref[lem:#1]{Lemma~\ref*{lem:#1}}}
\newcommand{\exref}[1]{\hyperref[ex:#1]{Example~\ref*{eg:#1}}}

\definecolor{deepcarminepink}{rgb}{0.94, 0.19, 0.22}
\definecolor{mediumelectricblue}{rgb}{0.01, 0.31, 0.59}
\definecolor{frenchblue}{rgb}{0.0, 0.45, 0.73}
\definecolor{green(munsell)}{rgb}{0.0, 0.66, 0.47}
\definecolor{violet(ryb)}{rgb}{0.53, 0.0, 0.69}
\definecolor{navyblue}{rgb}{0.0, 0.0, 0.5}
\definecolor{persimmon}{rgb}{0.93, 0.35, 0.0}
\definecolor{Plum}{rgb}{0.78, 0.08, 0.52}
\definecolor{ruby}{rgb}{0.88, 0.07, 0.37}

\definecolor{light-gray}{gray}{0.95}

\newcommand{\vqTxt}{variational query}

\newcommand{\vschTxt}{variational schema}

\newcommand{\tenvTxt}{type}

\newcommand{\vctxTxt}{variation context}

\newcommand{\vrelTxt}{variational relation}

\newcommand{\presCondTxt}{presence condition}

\newcommand{\fexpTxt}{feature expression}

\newcommand{\vCondTxt}{variational condition}

\newcommand{\edu}{\OB{\mathit{edu}}}

\newcommand{\vOne}{\OB{V_1}}
\newcommand{\vTwo}{\OB{V_2}}
\newcommand{\vThree}{\OB{V_3}}
\newcommand{\vFour}{\OB{V_4}}
\newcommand{\vFive}{\OB{V_5}}
\newcommand{\tOne}{\OB{T_1}}
\newcommand{\tTwo}{\OB{T_2}}
\newcommand{\tThree}{\OB{T_3}}
\newcommand{\tFour}{\OB{T_4}}
\newcommand{\tFive}{\OB{T_5}}
\newcommand{\fOne}{\OB{f_1}}
\newcommand{\fTwo}{\OB{f_2}}
\newcommand{\engemp}{\OB{\mathit{engineerpersonnel}}}
\newcommand{\othemp}{\OB{\mathit{otherpersonnel}}}

\newcommand{\empacct}{\OB{\mathit{empacct}}}
\newcommand{\empno}{\OB{\mathit{empno}}}
\newcommand{\name}{\OB{\mathit{name}}}
\newcommand{\hiredate}{\OB{\mathit{hiredate}}}
\newcommand{\titleatt}{\OB{\mathit{title}}}
\newcommand{\deptname}{\OB{\mathit{deptname}}}
\newcommand{\salary}{\OB{\mathit{salary}}}
\newcommand{\job}{\OB{\mathit{job}}}
\newcommand{\deptno}{\OB{\mathit{deptno}}}
\newcommand{\dept}{\OB{\mathit{dept}}}
\newcommand{\managerno}{\OB{\mathit{managerno}}}
\newcommand{\empbio}{\OB{\mathit{empbio}}}
\newcommand{\sex}{\OB{\mathit{sex}}}
\newcommand{\birthdate}{\OB{\mathit{birthdate}}}
\newcommand{\fname}{\OB{\mathit{firstname}}}
\newcommand{\lname}{\OB{\mathit{lastname}}}
\newcommand{\course}{\OB{\mathit{course}}}
\newcommand{\ecourse}{\OB{\mathit{ecourse}}}
\newcommand{\student}{\OB{\mathit{student}}}

\newcommand{\cname}{\OB{\mathit{coursename}}}
\newcommand{\tno}{\OB{\mathit{teacherno}}}
\newcommand{\sno}{\OB{\mathit{studentno}}}
\newcommand{\cno}{\OB{\mathit{courseno}}}
\newcommand{\class}{\OB{\mathit{class}}}
\newcommand{\timeatt}{\OB{\mathit{time}}}
\newcommand{\teach}{\OB{\mathit{teach}}}
\newcommand{\take}{\OB{\mathit{take}}}
\newcommand{\grade}{\OB{\mathit{grade}}}
\newcommand{\isstudent}{\OB{\mathit{std}}}
\newcommand{\isteacher}{\OB{\mathit{instr}}}
\newcommand{\studentnum}{\OB{\mathit{stdnum}}}
\newcommand{\teachernum}{\OB{\mathit{instrnum}}}

\newcommand{\rSep}{\rn{-}}

\newcommand*{\rE}{\rSep\rn{E}}

\newcommand*{\relation}{\rn{Relation}}
\newcommand*{\relationE}{\relation\rE}
\newcommand*{\choice}{\rn{Choice}}
\newcommand*{\choiceE}{\choice\rE}
\newcommand*{\product}{\rn{Product}}
\newcommand*{\productE}{\product\rE}
\newcommand*{\setop}{\rn{SetOp}}
\newcommand*{\setopE}{\setop\rE}
\newcommand*{\prj}{\rn{Project}}
\newcommand*{\prjE}{\prj\rE}
\newcommand*{\sel}{\rn{Select}}
\newcommand*{\selE}{\sel\rE}

\newcommand*{\rC}{\rSep\rn{C}}
\newcommand*{\bool}{\rn{Boolean}}
\newcommand*{\boolC}{\bool\rC}
\newcommand*{\nott}{\rn{Neg}}
\newcommand*{\notC}{\nott\rC}
\newcommand*{\conj}{\rn{Conjunction}}
\newcommand*{\conjC}{\conj\rC}
\newcommand*{\disj}{\rn{Disjunction}}
\newcommand*{\disjC}{\disj\rC}
\newcommand*{\attVal}{\rn{AttOptVal}}
\newcommand*{\attValC}{\attVal\rC}
\newcommand*{\attAtt}{\rn{AttOptAtt}}
\newcommand*{\attAttC}{\attAtt\rC}
\newcommand*{\choiceC}{\choice\rC}

\newcommand*{\fName}{\OB{f}}
\newcommand*{\fSet}{\OB{\mathbf{F}}}
\renewcommand*{\dimMeta}{\OB{e}}
\newcommand*{\fModel}{\OB{m}}
\newcommand*{\ffSet}{\OB{\mathbf{E}}}
\newcommand*{\bTag}{\OB{b}}
\newcommand*{\bSet}{\mathbf{B}}
\newcommand*{\config}{\OB{c}}
\newcommand*{\confSet}{\OB{\mathbf{C}}}
\newcommand*{\fct}[1][\config]{\OB{\dimMeta_{#1}^t}}
\newcommand*{\vctx}{\OB{\dimMeta}}

\newcommand*{\pDB}{\OB{\underline{\db}}}
\newcommand*{\pQ}{\OB{\underline{\q}}}
\newcommand*{\pQSet}{\OB{\underline{\qSet}}}
\newcommand*{\pAtt}{\OB{\underline{\att}}}
\newcommand*{\pAttList}{\OB{\underline{\attList}}}
\newcommand*{\pRel}{\OB{\underline{\rel}}}
\newcommand*{\pRelSch}{\OB{\underline{\relSch}}}
\newcommand*{\pRelCont}{\OB{\underline{\relCont}}}
\newcommand*{\pRelContSet}{\OB{\underline{\relContSet}}}
\newcommand*{\pSch}{\OB{\underline{\sch}}}
\newcommand*{\pInst}{\OB{\underline{\dbInst}}}
\newcommand*{\pInstSet}{\OB{\underline{\dbInstSet}}}
\newcommand*{\pAttSet}{\OB{\underline{\attSet}}}

\newcommand*{\pRelSchSet}{\OB{\underline{\relSchSet}}}
\newcommand*{\pSchSet}{\OB{\underline{\schSet}}}

\newcommand*{\pElem}{\OB{\underline \elem}}
\newcommand*{\pTuple}{\OB{\underline \tuple}}
\newcommand*{\pTab}{\OB{\underline \tab}}

\newcommand*{\db}{\OB{D}}
\newcommand*{\q}{\OB{q}}
\newcommand*{\att}{\OB{a}}
\newcommand*{\rel}{\OB{r}}
\newcommand*{\tab}{\OB{t}}
\newcommand*{\relSch}{\OB{s}}
\newcommand*{\relCont}{\OB{U}}
\newcommand*{\attList}{\OB{A}}
\newcommand*{\sch}{\OB{S}}
\newcommand*{\dbInst}{\OB{\mathcal{I}}}
\newcommand*{\tuple}{\OB{u}}
\newcommand*{\annot}[2][\dimMeta]{\OB{{#2}^{#1}}}
\newcommand*{\elem}{\OB{x}}

\newcommand*{\cond}{\OB{\theta}}
\newcommand*{\pset}{\OB{\underline {X}}}

\newcommand*{\qSet}{\OB{\mathbf{Q}}}
\newcommand*{\dbInstSet}{\OB{\mathbf{I}}}
\newcommand*{\attSet}{\OB{\mathbf{A}}}

\newcommand*{\schSet}{\OB{\mathbf{S}}}
\newcommand*{\relSchSet}{\OB{\mathbf{S}}}
\newcommand*{\relContSet}{\OB{\mathbf{R}}}
\newcommand*{\condSet}{\OB{\mathbf{\Theta}}}

\newcommand*{\vDB}{\db}
\newcommand*{\vSchDef}{\OB{\setDef {{\vRelSch_1 , \ldots, \vRelSch_n}}^\fModel}}
\newcommand*{\vdbInst}{\OB{\dbInst}}
\newcommand*{\vdbInstSet}{\OB{\dbInstSet}}

\newcommand*{\vQ}{\OB{\q}}
\newcommand*{\vRelSch}{\OB{\relSch}}
\newcommand*{\vRelCont}{\OB{\relCont}}
\newcommand*{\vSch}{\OB{\sch}}
\newcommand*{\vset}{\OB{X}}
\newcommand*{\vTab}{\tab}
\newcommand*{\vRel}{\rel}
\newcommand*{\vAtt}{\att}

\makeatletter
\def\vRelConfed{%
   \@ifnextchar[%
     {\vRelConfed@i}
     {\vRelConfed@i[\config]}%
}
\def\vRelConfed@i[#1]{%
   \@ifnextchar[%
     {\vRelConfed@ii{#1}}
     {\vRelConfed@ii{#1}[\pRel]}%
}
\def\vRelConfed@ii#1[#2]{%
\OB{{#2}^{#1}}
}
\makeatother

\makeatletter
\def\optAtt{%
   \@ifnextchar[%
     {\optAtt@i}
     {\optAtt@i[\dimMeta]}%
}
\def\optAtt@i[#1]{%
   \@ifnextchar[%
     {\optAtt@ii{#1}}
     {\optAtt@ii{#1}[\vAtt]}%
}
\def\optAtt@ii#1[#2]{%
\OB{\annot[#1]{#2}}
}
\makeatother

\newcommand*{\vAttList}{\OB{\attList}}

\makeatletter
\def\vRelDef{
   \@ifnextchar[
      {\vRelDef@i}
      {\vRelDef@i[\vRel]}
}
\def\vRelDef@i[#1]{
   \@ifnextchar[
      {\vRelDef@ii{#1}}
      {\vRelDef@ii{#1}[\dimMeta]}
}
\def\vRelDef@ii#1[#2]{
   \@ifnextchar[
      {\vRelDef@iii{#1}{#2}}
      {\vRelDef@iii{#1}{#2}[\vAttList]}
}
\def\vRelDef@iii#1#2[#3]{
\OB{{#1}\left({#3}\right)^{#2}}
}

\newcommand*{\vRelDefNum}[1]{\OB{\vRelDef [\vRel_{#1}] [\dimMeta_{#1}] [\vAttList_{#1}] }}

\newcommand*{\vAttSet}{\OB{\attSet}}

\newcommand*{\vRelSchSet}{\OB{\relSchSet}}
\newcommand*{\vRelContSet}{\OB{\relContSet}}
\newcommand*{\vSchSet}{\OB{\schSet}}
\newcommand*{\vInstSet}{\OB{\dbInstSet}}

\newcommand*{\emp}{\OB{\varepsilon}}
\newcommand*{\empAtt}{\OB{\varepsilon}}
\newcommand*{\empRel}{\OB{\varepsilon}}
\newcommand*{\vAttOpCte}{\OB{\op \pAtt \cte}}
\newcommand*{\vAttOpAtt}{\OB{\op {\pAtt_1} {\pAtt_2}}}
\renewcommand*{\tag}[2]{\OB{{#1}^{#2}}}
\newcommand*{\pc}{\OB{pc}}
\newcommand*{\getPC}[1]{\OB{pc\left(#1\right)}}
\newcommand*{\getRel}[1]{\OB{rel\left(#1\right)}}
\newcommand*{\getAtt}[1]{\OB{att\left(#1\right)}}

\newcommand*{\pCond}{\OB{\underline{\cond}}}
\newcommand*{\cte}{\OB{k}}
\newcommand*{\pCondSet}{\OB{\underline{\condSet}}}

\newcommand*{\vCond}{\OB{\cond}}
\newcommand*{\vCondSet}{\OB{\condSet}}

\newcommand*{\vType}{\OB{\vAttList}}

\newcommand{\myOR}{\hspace{1.5ex}|\hspace{1.5ex}}
\newcommand{\VVal}[1]{#1'}

\newcommand{\synDef}[2]{\OB{#1 \in #2 }}
\newcommand{\eqq}{\OB{\Coloneqq}}
\renewcommand{\t}{\OB{\prog{true}}}
\newcommand{\f}{\OB{\prog{false}}}

\newcommand*{\fS}{\OB{\mathbb{E}}}
\newcommand*{\fSem}[2][\config]{\OB{\fS\sem[#1]{#2}}}

\newcommand*{\elemS}{\OB{\mathbb{X}}}
\newcommand*{\elemSem}[2][\config]{\OB{\elemS\sem[#1]{#2}}}

\newcommand*{\elemG}{\OB{\mathcal{G}}}
\newcommand*{\elemGroup}[1][\elem]{\OB{\elemG\left({#1}\right)}}

\newcommand*{\qG}{\OB{\mathcal{Q}}}
\newcommand*{\qGroup}[1]{\OB{\qG\left({#1}\right)}}

\newcommand*{\cG}{\OB{\mathcal{C}}}
\newcommand*{\cGroup}[1][\vCond]{\OB{\cG\left({#1}\right)}}

\newcommand*{\aG}{\OB{\mathcal{A}}}
\newcommand*{\aGroup}[1][\vAttList]{\OB{\aG\left({#1}\right)}}

\newcommand*{\olS}{\OB{\mathbb{A}}}
\newcommand*{\olSem}[2][\config]{\OB{\olS\sem[#1]{#2}}}

\newcommand*{\orS}{\OB{\mathbb{R}}}
\newcommand*{\orSem}[2][\config]{\OB{\orS\sem[#1]{#2}}}

\newcommand*{\otS}{\OB{\mathbb{T}}}
\newcommand*{\otSem}[2][\config]{\OB{\otS\sem[#1]{#2}}}

\newcommand*{\ouS}{\OB{\mathbb{U}}}
\newcommand*{\ouSem}[2][\config]{\OB{\ouS\sem[#1]{#2}}}

\newcommand*{\ovS}{\OB{\mathbb{V}}}
\newcommand*{\ovSem}[2][\config]{\OB{\ovS\sem[#1]{#2}}}

\newcommand*{\osS}{\OB{\mathbb{S}}}
\newcommand*{\osSem}[2][\config]{\OB{\osS\sem[#1]{#2}}}

\newcommand*{\odbS}{\OB{\mathbb{I}}}
\newcommand*{\odbSem}[2][\config]{\OB{\odbS\sem[#1]{#2}}}

\newcommand*{\ecS}{\OB{\mathbb{C}}}
\newcommand*{\ecSem}[2][\config]{\OB{\ecS\sem[#1]{#2}}}

\newcommand*{\eeS}{\OB{\mathbb{Q}}}
\newcommand*{\eeSem}[2][\config]{\OB{\eeS\sem[#1]{#2}}}

\newcommand{\setDef}[1]{\OB{\{#1\}}}

\newcommand{\op}[2]{\OB{#1\bullet#2}}
\newcommand{\annd}[1]{\OB{{#1}_1 \wedge {#1}_2}}
\newcommand{\orr}[1]{\OB{{#1}_1 \vee {#1}_2}} 

\newcommand{\vPrj}[2][\vAttList]{\OB{\pi_{#1} #2}}
\newcommand{\vSel}[2][\vCond]{\OB{\sigma_{#1} #2}}

\makeatletter
\def\env{
  \@ifnextchar[
    {\env@i}
    {\env@i[\vctx]}
}
\def\env@i[#1]{
  \@ifnextchar[
    {\env@ii{#1}}
    {\env@ii{#1}[\vSch]}
}
\def\env@ii#1[#2]#3#4{
{\OB{#1,#2 \vdash#3:#4}}
}
\makeatother

\newcommand{\envWithSchema}[2][\OB{\vctx}]{\env[#1]{\vRel}{#2}}

\newcommand{\envOne}[1][\vctx]{\env[#1]{\vQ_1}{\envInContext [\vctx_1] \vType_1}}
\newcommand{\envTwo}[1][\vctx]{\env[#1]{\vQ_2}{\envInContext [\vctx_2] \vType_2}}
\newcommand{\envPrime}{\env{\vQ}{\envInContext[\VVal \vctx]{\VVal \vType}}}

\newcommand{\envCond}[2][\vctx, \vType]{\OB{#1\vdash #2}}
\newcommand{\envCondAnnot}[2][\vctx, {\annot [\VVal \vctx] \vType}]{\OB{#1\vdash #2}}

\newcommand{\envInContext}[2][\vctx]{\OB{{#2}^{#1}}}
\newcommand{\envEval}[2]{#1 \equiv #2}
\newcommand{\imply}{\OB{\rightarrow}}
\newcommand{\subsume}[2]{\OB{#1 \prec #2}}
\newcommand{\nsubsume}[2]{\OB{#1 \not \prec #2}}

\newcommand{\sat}[1]{\OB{\mathit{sat}\left(#1\right)}}
\newcommand{\taut}[1]{\OB{\mathit{taut}\left(#1\right)}}

\renewcommand{\dom}[2][\vdbInst]{\OB{\mathit{dom}_{#1}\left(#2\right)}}
\newcommand{\type}[1][\pAtt]{\OB{\mathit{type}\left(#1\right)}}

\newcommand{\spcEq}[1][0cm]{\OB{\hspace{#1}=}}

\newcommand{\paran}[1]{\OB{\left(#1\right)}}

\begin{document}

\title{Managing Variability in Relational Databases by VDBMS }

\author{Parisa Ataei, Qiaoran Li, Eric Walkingshaw, Arash Termehchy}
\email{[ataeip, liqiao, walkiner, termehca]@oregonstate.edu}
\affiliation{%
  \institution{Oregon State University}
  \city{Corvallis}
  \state{Oregon}
}





\begin{abstract}
Variability inherently exists in databases in various contexts which creates 
database variants.
For example,
variants of a database could have different schemas/content (database 
evolution problem), variants of a database could root from 
different sources (data integration problem), variants of a database
could be deployed differently for specific application domain (deploying
a database for different configurations of a software system), etc. 
Unfortunately, while there are specific solutions to each of the 
problems arising in these contexts, there is no general solution
that accounts for variability in databases and addresses managing 
variability within a database.
In this paper, we formally define \emph{variational databases (VDBs)} and 
statically-typed 
\emph{variational relational algebra (VRA)} to query VDBs---both 
database and queries explicitly account for variation.
We also design and implement \emph{variational database management
system (VDBMS)} to run variational queries over a VDB effectively and efficiently.
To assess this, we generate two VDBs from real-world databases in the context of software
development and database evolution with a set of experimental queries for each. 
\end{abstract}



\keywords{Variational Database, Variational Relational Algebra, 
Managing Data Variability}


\maketitle

\section{Introduction}
\label{sec:intro}

Data and variants of data are abundant. 
Having variants of data is unavoidable, it needs to be maintained, 
and it appears in multiple context.
While there are solutions dealing with variants of data in specific contexts,
there is no general solution to the problem of accounting for variability in data.
In this section, we focus on variability in relational databases in different contexts 
that already exists in the field,
current solutions that address them, and their shortcomings.

\textbf{Context 1: schema evolution.}
One way that data variants arise is when a database schema evolves over 
time--- an unpredictable and inevitable occurrence which is 
independent from how good or bad
the initial schema has been designed~\cite{dbDecay16Stonebraker}.
Although, there are guidelines that direct
database administrators (DBAs) to design schemas that do not rapidly decay over time, 
in practice, there is still no way to avoid the change due to 
business circumstances, mergers and acquisitions, and 
new applications for the same data~\cite{dbDecay16Stonebraker}. 
\emph{Every time the database evolves, a new ``variant'' is created.}

Current solutions addressing schema evolution 
require DBAs to design a unified schema, map old schema
to the unified one, and migrate the database to keep the old variants of 
database~\cite{prima08Moon}. 
They also require DBAs (and/or application developers) to
rewrite queries which were based on the previous schema.
These approaches have multiple limitations:
1) they are time-consuming, error-prone, and burdensome tasks for DBAs and developers
and
2) users cannot access any of the old variants even if they desire. 
Some of the recent solutions address these issues to some extent, but
induce other limitations on the problem, explored more in \secref{mot}.

\textbf{Context 2: data integration.}
Another kind of common variation in data arises from the
integration of multiple different data sources into a single database,
i.e., data integration~\cite{dataIntegBook}. Each data source is a data variant.
Assuming that all sources have the
same format (e.g., they are all relational databases), they still could differ in details such as 
their schema. 

Existing solutions for integrating all variants requires the capability to adjust all databases
to a unified database. For example, in the case of relational databases, we need to map
all schemas to a unified schema~\cite{dataIntegBook}. Here, the variants
of data come from different sources that need to be combined systematically.
The problem with unifying all the variants into a unified database
is that 
queries cannot specify a unique source to inquire information from, the same
problem appeares in schema evolution.
Furthermore, in this context, a user's inquiry result is collected from all variants
and presented to the user. Unfortunately, there is no way for the user
to identify which variant a specific piece of result (e.g., a tuple in the case of relational databases)
belongs to, i.e., data integration systems lose data provenance~\cite{dataIntegBook}.

\textbf{Context 3: Software Product Lines.}
Additionally, data variants arises as the artifact of software development, evolution, and 
maintenance, especially in the context of software product line (SPL).
SPL is an
approach to developing and maintaining software-intensive systems in a cost-effective,
easy to maintain manner. 
The products of a SPL pertain to a
common application domain or business goal. They also have a common
managed set of features that describe the specific need for a product. They share 
a common codebase which is used to produce a product with respect to its set of 
selected (enabled) features~\cite{splBook}. 
Different products of a SPL typically have different
sets of features or are tailored to run in different environments. These
differences impose different data requirements. For example, different legal
requirements often require tracking different data in products tailored for use
in different countries or regions.

In practice, software systems produced by a SPL are accommodated with a database that
has all attributes and tables available in all variants-- a global database~\cite{vdbSpl18ATW}. 
Unfortunately, this approach is
inefficient, error-prone, and filled with lots of null values since not all attributes and tables
are valid for all variant products.

As shown thus far, variability in databases is abundant, inevitable, inexorable,
and impacts DBAs, data scientists, and developers significantly.
Various research addresses \emph{instances}
of introduction/existence of variability into/in a database such as schema evolution,
although these customized solutions have some shortcomings within and beyond their own
context.
Current approaches are limited because they dismiss that:
\emph{the problem at hand
is essentially an instance of introduction/existence of variability into/in a database.} 

To solve this problem, we need to resolve the challenge of
how to incorporate variability in databases to allow for
expressive and efficient queries that satisfy the general needs of different contexts where variability 
appears? Our contributions in this paper address this challenge:
\begin{itemize}
\item We provide a framework to capture variability within a database using
propositional formulas over  
sets of features, called \emph{feature expressions}, following~\cite{vdb17ATW}.
\item We incorporate feature expressions  into both the content (tuples) and structure (schema)
of the database, introducing \emph{variational schemas}, \secref{vsch}, 
and \emph{variational tables}, \secref{vtab}, and together \emph{VDBs},
\secref{vdb}.
\item To express user information needs we define a domain-specific language
which is a combination of relational algebra and 
choice calculus~\cite{EW11tosem,Walk13thesis}, \secref{vrel-alg}, that
allows users to query a VDB by a \emph{variational query (v-query)}, \secref{vq}.
\item We define a static type system for our language, \secref{type-sys},
along with rules to minimize variation within a v-query, \secref{var-min}.
\item To efficiently query a 
variational database and receive effective clear results
we introduce \emph{Variational Database Management System (VDBMS)}, \secref{impl}.
\end{itemize}

To evaluate VDBMS, we adopt two real-world databases 
(employee database\footnote{\url{https://github.com/datacharmer/test_db}} 
and Enron email data corpus\footnote{\url{http://www.ahschulz.de/enron-email-data/}}) 
from two different contexts (schema evolution and SPL),
generate their counterpart VDBs (based on the changes applied to them), assess
the performance of VDBMS using a set of reasonable queries within each context, and
finally analyze and discuss the results, \secref{exp-disc}.
\section{Motivating Example}
\label{sec:mot}

\begin{table*}[t]
\caption{Schema evolution of an employee database for a SPL.
A feature (a boolean variable) represents an optional
inclusion of a schema or elements of a schema.  
The dash-underlined attributes in left schema column only exist when the education feature of SPL is enabled. This table represents 1056 possible schema variants 
(\ensuremath{2^5} schema variants when \edu\ is disabled in 
addition to \ensuremath{2^5 \times 2^5} when \edu\ is enabled).}
\label{tab:mot}
\begin{center}
\small
\begin{tabular} {| c | l || l | c |}
\hline
\textbf{Temporal} & \multicolumn{2}{ c |}{\textbf{SPL Features}} & \textbf{Temporal}\\
\cline{2-3}
\textbf{Features} & \multicolumn{1}{ c ||} {$\edu \vee \neg \edu \equiv \t$} & \multicolumn{1}{ c |} {\edu} & \textbf{Features}\\
\hline
\multirow{3}{*}{\vOne} &  \engemp\ (\empno, \name, \hiredate,\titleatt,\deptname) & 
\course\ (\cname, \tno) & \multirow{3}{*}{\tOne}\\
& \othemp\ (\empno, \name, \hiredate, \title, \deptname)  & \student\ (\sno, \cname) &\\
& \job\ (\titleatt, \salary) &  &\\
\hline
\multirow{2}{*}{\vTwo} & \empacct\ (\empno, \name, \hiredate, \titleatt, \deptname) & \course\ (\cno, \cname, \tno) & \multirow{2}{*}{\tTwo}\\
& \job\ (\titleatt, \salary) & \student\ (\sno, \cno) & \\
\hline
\multirow{4}{*}{\vThree} & \empacct\ (\empno, \name, \hiredate, \titleatt, \deptno) & \course\ (\cno, \cname) & \multirow{4}{*}{\tThree}\\
& \job\ (\titleatt, \salary) & \teach\ (\tno, \cno) &\\
& \dept\ (\deptname, \deptno, \managerno) & \student\ (\sno, \cno, \grade) &\\
& \empbio\ (\empno, \sex, \birthdate) & &\\
\hline
\multirow{4}{*}{\vFour} & \empacct\ (\empno, \hiredate, \titleatt, \deptno, \dashuline{\isstudent}, \dashuline{\isteacher}) & \ecourse\ (\cno, \cname) & \multirow{4}{*}{\tFour}\\
& \job\ (\titleatt, \salary) & \course\ (\cno, \cname, \timeatt, \class) & \\
& \dept\ (\deptname, \deptno, \managerno) & \teach\ (\tno, \cno) & \\
& \empbio\ (\empno, \sex, \birthdate, \name) & \student\ (\sno, \cno, \grade) & \\
\hline
\multirow{4}{*}{\vFive} & \empacct\ (\empno, \hiredate, \titleatt, \deptno,  \dashuline{\isstudent}, \dashuline{\isteacher}, \salary) & \ecourse\ (\cno, \cname, \deptno) & \multirow{4}{*}{\tFive}\\
& \dept\ (\deptname, \deptno, \managerno,  \dashuline{\studentnum}, \dashuline{\teachernum}) & \course\ (\cno, \cname, \timeatt, \class, \deptno) & \\
& \empbio\ (\empno, \sex, \birthdate, \fname, \lname) & \teach\ (\tno, \cno) & \\
&& \take\ (\sno, \cno, \grade) & \\
\hline
\end{tabular}
\end{center}
\end{table*}

Encoding variability into databases is more critical when multiple contexts meet. 
An existing example of this is when
the evolution of a SPL results in its database schema evolution~\cite{dbSPLevolve}. 
Database evolution within a SPL happens at two stages: 
1) \emph{component evolution}: 
developers update, refactor, improve, and components resulting in potentially 
changing the data model, i.e., the database schema,
and
2) \emph{product evolution}: 
clients want to add/remove a feature or component which potentially propagates
to database evolution~\cite{dbSPLevolve,splEvolveBP14}.

Products of a SPL require different variants of a database
to store their information, if any, since each product desires 
to store some specific 
information that others may not care about. 
These database variants differ mainly 
in their schema, a relation/attribute can either be included or excluded for a 
specific software product and the feature set for a product determines which
relations and attributes to include in its database, i.e., the features dictate the database
schema~\cite{vdbSpl18ATW}. 
In practice, SPLs use a unified schema for all database variants---an error-prone
approach resulting in lots of null values for attributes that do not exists in a 
database variant. 


%
We illustrate an example of
database schema
evolution within a SPL in an employee database.
We use parts
of this example as our running example throughout the paper.\footnote{The example is borrowed 
from~\cite{prima08Moon} with some adaptation and addition.}
\tabref{mot} outlines schema variations of our example. 
The schema consists of two sub-schemas, the left one (the second column from left)
stores employees information and the right one stores education-related information 
of employees while they are working at the company.
It has three dimensions of variability.
The horizontal axis accounts for SPL features---for brevity, 
we consider an SPL that has only one feature: \edu, 
if the \edu\ feature is enabled the database schema includes the education
sub-schema in addition to the employee sub-schema, otherwise it only consists
of the employee sub-schema.
The vertical axes on the left and right account for the other two dimensions, 
representing temporal changes to the employee and education  sub-schemas,
respectively.
The sub-schemas evolve over time due to business requirements~\cite{prima08Moon},
potentially with different paces which requires two different dimensions of variability 
to capture each of them individually. We use \emph{temporal features} to refer 
to each time a schema evolves. 
Enabling a temporal feature indicates that its corresponding database variants
include the sub-schema associated with the feature.
E.g., if \vOne\ is enabled the schema includes \engemp, \othemp, and \job.

We describe the schema variations shown in \tabref{mot} w.r.t.
SPL component evolution (when developers 
introduce new changes).
The database initially has three tables for all SPL products: 
\engemp, \othemp, and \job. In addition to these it has two tables \course\ and
\student\ for SPL products providing education. Note that teachers and 
students are both assumed to be employees.

To uniformly manage employee information, in variants of \vTwo, 
SPL developers decide to
combine \engemp\ and \othemp\ into one table \empacct. At 
the same time, in variants of \tTwo, 
due to increase in number of courses offered the \cno\ attribute
is added to \course\ table.

As companies grow, in temporal variants of \vThree, 
the \dept\ and \empbio\ tables are added. 
Similarly for variants of \tThree,
the \teach\ table is added in addition to the \grade\ attribute
in the \student\ table to account for employees attempt to 
better themselves.

Within \vFour\ schema variants, the \name\ attribute is moved from 
the \empacct\ table to the \empbio\ table.
Also, due to an increase of the number of employees who take courses,
attributes \isstudent\ and \isteacher\ are added to \empacct\ to account
for an employee being a student and/or teacher, respectively. Note that
these attributes only exists in a schema if the company offers education
to its employees.
In temporal variants of \tFour,
due to the high requests of employees, e-courses are 
offered, hence, the \ecourse\ table is added. Note that we could have 
a different design and consider offering e-courses as a SPL feature.

Eventually, to motivate employees the companies of temporal variants
\vFive\ consider salary dependent on the employee and not the title. Hence,
the \salary\ attribute is added to the \empacct\ table and the \job\ table is 
dropped.
To access statistics of departments for educational purpose, attributes
\studentnum\ and \teachernum\ are added to \dept\ relation when \edu\
is enabled.
Additionally, the first name and last name of employees is stored
separately and the \student\ table is renamed to \take. 

The schema also evolves due to product evolution, i.e.,
a client requests a new version. For example, consider a client  that
previously has requested software with database variant associated 
with \vFive.
Now the client decides to educate its employees.
Hence, the SPL developers need to reconfigure the product for such a
client to meet their needs. The schema evolution described above also captures
product evolution.

As mentioned earlier, existing solutions burden DBAs heavily by requiring
them to define a unified schema, map old schemas to the unified one, 
migrate the data, and rewrite queries~\cite{Prism08Curino}. 
Some approaches automate schema mapping when the schema evolves
to migrate the data, 
however, they still require the DBA to rewrite 
queries~\cite{mappingSchEvol03Velegrakis,semanticAdaptationSchEvol05Yu}.
Other solutions, keep some sort of history of database evolution and either allow the
user to specify the version of the database their query is written again and generate
queries for all other versions~\cite{prima08Moon}.
The specific problem of schema evolution when a SPL evolves is addressed by
designing
a new domain-specific language so that SPL developers 
can write scripts of the schema changes~\cite{dbSPLevolve},
which still requires a great effort by DBAs and SPL developers.
This amount of work grows exponentially 
as the number of potential variants grow, which in our example depends on the temporal
changes and SPL features. As a reminder, a SPL usually has hundreds of 
features~\cite{cppSpl}. As the SPL and its database evolve,
manually managing the variants becomes impossible. 

Although current approaches address variability in specific applications 
and they provide interesting efficient solutions, they are tailored to 
the underlying application and thus fail to address the more general problem.
This results in a pressing need of 
incorporating variability within databases in a flexible manner that can be applied in
various context which we address by introducing variational databases 
and VDBMS to interact with them.

\section{Preliminaries}
\label{sec:prelim}

In this section, we introduce concepts and notations that we use
throughout the paper. \tabref{notations} 
provides a short overview 
and is meant as an aid to find definitions
faster.
Throughout the paper, we discuss relational concepts and their
variational counterparts. For clarity, when we need to emphasize 
an entity is not variational we underline it, e.g., \pElem\ is a 
non-variational entity while \elem\ is its potentially counterpart.

\begin{table}
\caption{Introduced notations and terminologies with their corresponding section(s).}
\label{tab:notations}
\begin{center}
\small
\begin{tabular}{ |l c r| } 
 \hline
Name & Notation & \S \\
\hline
Feature & \fName & \multirow{7}{*} {\secref{encode-var}} \\
Feature expression & \dimMeta & \\
Annotated element \elem\ by \dimMeta & \annot \elem & \\
Configuration & \config & \\
True feature set of \config & \fct & \\
Evaluation of \dimMeta\ under \config & \fSem \dimMeta & \\
Presence condition of entity \elem & \getPC \elem & \\
\hline
Optional attribute & \vAtt & \multirow{4}{*}{\secref{vsch}}\\
Variational attribute set & \vAttList & \\
Variational relation schema & \vRelSch & \\
Variational schema & \vSch & \\
\hline
Variational tuple & \tuple & \multirow{3}{*}{\secref{vtab}} \\
Variational relation content & \vRelCont & \\
Variational table & \vTab = (\vRelSch, \vRelCont) & \\
\hline
Choice & $\chc {x, y}$ & \multirow{3}{*}{\secref{vrel-alg}}\\
Variational condition & \vCond & \\
Variational query & \vQ & \\
\hline
Configure query \vQ\ with configuration \config & \eeSem \vQ & \multirow{2}{*}{\secref{vra-sem}}\\
Group query \vQ & \qGroup \vQ &\\
\hline
\end{tabular}
\end{center}
\end{table}

\subsection{Relational Databases}
\label{sec:rel-db}

A relational database \pDB\ stores information in a structured manner by forcing
data to conform to a \emph{schema} \pSch\ that is a finite set 
$\setDef {{\pRelSch}_1, \ldots, {\pRelSch}_n}$ of \emph{relation schemas}.
A relation schema is defined as
$\pRelSch = \pRel \paran {\vi \pAtt k}$ where each $\pAtt_i$ is an
\emph{attribute} of the relation \pRel. 
The function $\getRel {\pAtt}$ returns the relation of an attribute.
The function \type [\pAtt]\ returns the \emph{type} of attribute \pAtt.

The content of database \pDB\ is stored in the form of \emph{tuples}. A tuple \pTuple\
is a mapping between a list of relation schema attributes and their values, i.e.,
$\pTuple = \paran {\vi {\underline v} k}$ for the relation schema \pRel \paran {\vi \pAtt k}.
Hence a \emph{relation content}, \pRelCont, is a set of tuples \setDef {\vi \pTuple m}.
The function $\getAtt {\underline v}$ returns the attribute of a value.
A \emph{table} \pTab is a pair of relation content and relation schema.
A \emph{database instance}, \pInst, of the database \pDB\ with the
schema \pSch, is a set of relation contents 
$\setDef  {{\pRelCont}_1, \ldots, {\pRelCont}_n}$ corresponding
to a set of relation schemas $\setDef {{\pRelSch}_1, \ldots, {\pRelSch}_n}$ 
defined in \pSch. 
For brevity, when it is clear from the context we refer to a database instance
by \emph{database}.

\subsection{Encoding Variability}
\label{sec:encode-var}

The first challenge of incorporating variability into a database
is to represent variability. To represent
variability we require a \emph{set of features}, denoted by \fSet, 
appropriate for the context that the database is used for.
For example, for the context of schema evolution, features can be generated from version 
numbers (e.g., features \vOne\ to \vFive\ and \tOne\ to \tFive\ in the 
motivating example, \tabref{mot}); for SPLs, 
the features can be adopted from the SPL feature set (e.g.: the \edu\ feature in
our motivating example, \tabref{mot}); and 
for data integration, the features can demonstrate resources.  
For simplicity and without loss of generality, 
features are assumed to be boolean variables, although, it is easy to extend them
to multi-valued variables. 

Assuming that all the features by default are set to \prog{false},
enabling some of them specifies a variant. Hence, to specify a variant
we define a function, called a \emph{configuration}, 
that maps every feature in the feature set to a boolean value.
By definition, a configuration is a \emph{total} function,
i.e., it includes \emph{all} features defined in the feature set. 
For brevity, we represent a configuration \config\ with a set of enabled features
which represents a variant. E.g., the configuration \ensuremath{
\setDef {\vTwo,\tThree,\edu}
}
represents a database variant with employee sub-schema associated with \vTwo\
and education sub-schema associated with \tThree in \tabref{mot}.

Having defined a set of features, we also need to incorporate them into the database.
To represent features in the database, we construct propositional formulas of features 
(which are basically boolean variables),
describing the condition (circumstance) where one or more variant are valid,
i.e., assigning features to their values defined in variant's configuration and 
evaluating the propositional formula results in \prog{true}. For example,
the propositional formula $\neg \edu$ represents all variants of our
motivating example that do not 
have the education part of the schema, i.e., variant schemas of the 
left schema column. 

We call propositional formula of features \emph{feature expression} and define
it formally in \figref{fexp-def}. 
We define the syntax of feature expressions in \figref{fexp-def}.
Feature expression semantics, denoted by 
$\fSem \dimMeta : \ffSet \to \confSet \to \bSet$,
evaluates feature expression \dimMeta\ under configuration \config,
also called configuration of feature expression \dimMeta\ under \config. 
For example, $\fSem [\{A\}] {A \vee B} = \t$
\figref{fexp-def} defines the syntax and equivalence of two feature expressions.
However, we define the evaluation of feature expressions and functions over them
in \appref{fexp}.
We define two functions over feature expressions, as shown in \figref{fexp-def}:
1) \emph{satisfiability}: feature expression \dimMeta\ is \emph{satisfiable} if there 
exists configuration \config\ s.t. \fSem \dimMeta = \t\
and 2) \emph{tautology}: feature expression \dimMeta\ is a \emph{tautology} if  
for all valid configurations we have: \fSem \dimMeta = \t.

To incorporate feature expressions into the database,
we \emph{annotate/tag} database elements (including attributes, relations, and tuples) 
with feature expressions. An \emph{annotated element} \elem\ with feature expression \dimMeta\
is denoted by \annot \elem\ where conceptually, feature expression \dimMeta\ represents
a group of configuration where their variants contain element \elem. 
The feature expression attached to an element is called a \emph{presence condition}
since it determines the condition under which the element is valid.
The function \getPC \elem\ returns the presence condition of the element
\elem.
For example, the
annotated number $2^A$ indicates that the number 2 is valid only when feature $A$ is 
enabled. Here, $\getPC {2} = A$.

No matter the context, features often times have a relationship with each other that
constrains configurations. For example, only one of the temporal features of \vOne -- \vFive\
can be \t\ for a given variant.
This relationship can easily be captured by a feature expression, called a \emph{feature model},
which restricts the set of valid configurations, i.e., 
if configuration \config\ violates the relationship then evaluating the feature model \fModel\
under this configuration evaluates to \f: \fSem \fModel = \f.
For example, the restriction that at a given time only one employee schema 
is valid is represented by the feature expression:
$\left(\vOne \wedge \neg \left(\vTwo \vee \hdots \vee \vFive \right) \right)
\vee \left(\vTwo \wedge \neg \left(\vOne \vee \vThree \vee \vFour \vee \vFive \right) \right) 
\vee \hdots 
\vee \left(\vFive \wedge \neg \left(\vOne \vee \hdots \vee \vFour \right) \right)$.
Note that this is not the feature model for the entire motivating example.

\begin{figure}
\textbf{Feature expression generic object:}
\begin{syntax}
\synDef \fName \fSet &\textit{Feature Name}
\end{syntax}

\medskip
\textbf{Feature expression syntax:}
\begin{syntax}
\synDef \bTag \bSet &\eqq& \t \myOR \f & \textit{Boolean Value}\\
\synDef \dimMeta \ffSet &\eqq& \bTag \myOR \fName \myOR \neg \fName \myOR \dimMeta \wedge \dimMeta \myOR \dimMeta \vee \dimMeta & \textit{Feature Expression}\\
\synDef \config \confSet &=& \fSet \to \bSet &\textit{Configuration}
\end{syntax}

\medskip
\textbf{Feature expression equivalence:}
\begin{alignat*}{1}
\dimMeta_1 \equiv \dimMeta_2 \textit{ iff \ } \forall \config \in \confSet: \fSem {\dimMeta_1} = \fSem {\dimMeta_2}
\end{alignat*}

\caption{Feature expression syntax and equivalence.}
\label{fig:fexp-def}
\end{figure}

\subsection{Variational Set}
\label{sec:vlist-vset}

%
A \emph{variational set (v-set)} $\vset = \setDef {\annot [\dimMeta_1] {\elem_1},\ldots, \annot [\dimMeta_n] {\elem_n}}$ 
is a set of annotated elements where the presence condition of elements is satisfiable~\cite{EWC13fosd,Walk14onward,vdb17ATW}.
%
%
%
Conceptually, a \emph{variational set} represents many different plain sets
that can be generated by enabling or disabling features
and including only the elements whose feature expressions evaluate to \t,
i.e., a variational set is a function from a configuration of its
features to the corresponding plain set. 
A plain set is denoted by \pset\ while a variational set of non-variational elements
is denoted by \ensuremath{\ddot \pset}.
We typically omit the feature
expression \prog{true} in a variational set,
e.g., the v-set 
$\setDef {\annot [\fName_1] 2, \annot [\fName_2] 3, 4}$
represents four plain sets under different configurations: 
\setDef {2,3,4} when $\fName_1$ and $\fName_2$
are enabled, \setDef {2,4} when $\fName_1$ is enable but $\fName_2$ is disabled,
\setDef {3,4} when $\fName_2$ is enabled but $\fName_1$ is disabled,
and \setDef {4} when both $\fName_1$ and $\fName_2$ are disabled.
%

A variational set itself can also be annotated with a feature expression.
An \emph{annotated variational set} $\annot \vset = \setDef {\annot [\dimMeta_1] {\elem_1},\ldots,\annot [\dimMeta_n] {\elem_n}}^\dimMeta$ is a
v-set that it is annotated itself by a \emph{feature expression} \dimMeta.
In essence, annotating a v-set with the feature expression \dimMeta\ 
restricts the condition under which its elements are valid, i.e., it forces
elements' presence conditions to be more specific. This restriction
can be captured by the property:
$\setDef {\annot [\dimMeta_1] {\elem_1} ,\ldots, \annot [\dimMeta_n] {\elem_n}}^\dimMeta
\equiv 
\setDef {\annot [\dimMeta_1 \wedge \dimMeta] {\elem_1},\ldots, \annot [\dimMeta_n \wedge \dimMeta] {\elem_n}}
$.
%
%
%
For example, the annotated variational set
$\{\annot [\fName_1] 2, \annot [\neg \fName_2] 3, 4, \annot [\fName_3] 5\}^{\fName_1 \wedge \fName_2}$
indicates that all the elements of the set can only exists
when both $\fName_1$ and $\fName_2$ are enabled. Thus, it is equivalent to
$\{\annot [\fName_1 \wedge \fName_2] 2,\annot [\fName_1 \wedge \fName_2] 4,\annot [\fName_1 \wedge \fName_2 \wedge \fName_3] 5\}$. The element $3$ is dropped from the set 
since its presence condition is unsatisfiable, i.e., $\neg \sat {\neg \fName_2 \wedge (\fName_1 \wedge \fName_2)}$.


We provide operation definitions over v-sets. These operations are vastly used
in \secref{type-sys}.

\begin{definition}[V-set union]
\label{def:vset-union}
The \emph {union} of two v-sets is the union of their elements with the disjunction of 
presence conditions if an element exists in both v-sets:
\ensuremath{
\vset_1 \cup \vset_2 = \setDef {\annot [\dimMeta_1] \elem \myOR \annot [\dimMeta_1] \elem \in \vset_1, \annot [\dimMeta_2] \elem \not \in \vset_2}
\cup \setDef {\annot [\dimMeta_2] \elem \myOR \annot [\dimMeta_2] \elem \in \vset_2, \annot [\dimMeta_1] \elem  \not \in \vset_1}
\cup \setDef {\annot [\dimMeta_1 \vee \dimMeta_2] \elem \myOR 
\annot [\dimMeta_1] \elem \in \vset_1, \annot [\dimMeta_2] \elem \in \vset_2}
}.
E.g., 
\ensuremath{
\setDef {2,\annot [\dimMeta_1] 3, \annot [\dimMeta_1] 4} \cup \setDef {\annot [\dimMeta_2] 3, \annot [\neg \dimMeta_1] 4} = \setDef {2, \annot [\dimMeta_1 \vee \dimMeta_2] 3, 4}
}.
\end{definition}

%

%
\begin{definition}[V-set intersection]
\label{def:vset-intersect}
The \emph{intersection} of two v-sets is a v-set of their shared elements
annotated with the conjunction of their presence conditions, i.e., 
\ensuremath{
\vset_1 \cap \vset_2 = \setDef {
\annot [\dimMeta_1 \wedge \dimMeta_2 ]\elem \myOR
\annot [\dimMeta_1] \elem \in \vset_1, \annot [\dimMeta_2] \elem \in \vset_2,
\sat {\dimMeta_1 \wedge \dimMeta_2}
}
}.
E.g., \ensuremath{
\setDef {2, \annot [\fName_1] 3, \annot [\neg \fName_2] 4} \cap
\annot [\fName_2] {\setDef{2,3,4,5}} =
\setDef{\annot [\fName_2] 2, \annot [\fName_1 \wedge \fName_2] 3}
}.
\end{definition}

\begin{definition} [V-set equivalence]
\label{def:vset-eq}
Two v-sets are \emph{equivalent}, denoted by
\ensuremath{\vset_1 \equiv \vset_2}, iff
\ensuremath{
\forall \annot [\dimMeta_1] \elem \in (\vset_1 \cup \vset_2). 
\annot [\dimMeta_1] \elem \in \vset_1, \annot [\dimMeta_2] \elem \in \vset_2, 
\dimMeta_1 \equiv \dimMeta_2},
i.e., they both cover the same set of elements and the presence conditions
of elements from the two v-sets are equivalent.
\end{definition}

\begin{definition} [V-set subsumption]
\label{def:vset-subsumption}
The v-set \ensuremath{\vset_1} \emph {subsumes} the v-set
\ensuremath{\vset_2}, $\subsume {\vset_2} {\vset_1}$, iff
\ensuremath{ \forall \annot [\dimMeta_2] \elem \in \vset_2.
\annot [\dimMeta_1] \elem \in \vset_1, 
\sat {\dimMeta_1 \wedge \dimMeta_2}
},
i.e., all elements in $\vset_2$ also exist in $\vset_1$ and 
s.t. the element is valid in a shared configuration between the v-sets.
E.g., 
\ensuremath{
 \subsume {\annot [\fName_1] {\setDef {2,3}}} {\setDef {2, \annot [\fName_1 \vee \fName_2] 3, 4}}},
however, 
\ensuremath{
 \nsubsume {\annot [\fName_1] {\setDef {2,3}}} {\setDef {2, \annot [\neg \fName_1 \wedge \fName_2] 3}}}
and
\ensuremath{
\nsubsume {\setDef {\annot [\fName_1] 2,\annot [\fName_1] 3, 4}} {\setDef {2, \annot [\fName_1 \wedge \fName_2] 3}}}.
\end{definition}

\section{Variational Database}
\label{sec:vdb}

A \emph{Variational Database (VDB)} is intuitively meaningful when a set of database 
instances with variations in the schema and/or content exists and a user's information
need requires accessing some or all of them simultaneously. 
To incorporate variability within a database, we annotate elements with feature expressions,
as introduced in \secref{encode-var}. We use annotated elements at two levels of a 
database: 1) schema, within a schema we allow attributes and relations to be valid 
conditionally based on the feature expression assigned to them, explicated in \secref{vsch}
and 2) content, we annotate each tuple with a feature expression, indicating when the tuple 
is valid, elucidated in \secref{vtab}. 
We also annotate the entire schema and table with a feature expression, 
resulting in a hierarchal structure of feature expressions, explored in 
\secref{vsch} and \secref{vtab}.

\subsection{Variational Schema}
\label{sec:vsch}

Variability can exist in the structure of data, i.e., the schema.
As motivated in \secref{mot}, schema variations include/exclude a/an relation/attribute. 
We annotate attributes, relations, and the schema itself with feature expressions,
making them exist variationally (conditionally).
A \emph{\vschTxt\ (v-schema)} is an annotated set of variational relation 
schemas:
$\synDef \vSch \vSchSet \eqq \vSchDef$.
The presence condition of \vschTxt, \fModel, determines all valid 
configurations for schema variants, however, it also captures the relationship
between features of the underlying application and their constraints,
as explained in \secref{encode-var} and introduced as the VDB feature model. 
Hence, the v-schema defines all valid schema variants of a VDB. 

A \emph{\vrelTxt\ (v-relation) schema}, \vRelSch, is a relation name
accompanied with an annotated variational attribute set:
$\synDef {\vRelSch} \vRelSchSet \eqq \vRelDef$.
The presence condition of the relation schema, \dimMeta, determines the
set of all possible relation schema variants for relation \pRel.
A \emph{variational attribute set}, \vAttList, is a variational set of attributes,
i.e.,
$\synDef \vAttList \vAttSet \eqq \optAtt, \vAttList \myOR \empAtt$,
where \empAtt\ denotes an empty attribute. \exref{vsch} illustrates creating a v-schema.

\begin{example}
\label{eg:vsch}
The v-schema of a VDB including only relations \empacct\ and \ecourse\ in the last two rows
of \tabref{mot}, where valid features are \vFour, \vFive, \edu, \tFour, \tFive, is:
%
\begin{align*}
\vSch_1 &=
\{ \empacct ( \empno, \hiredate, \titleatt, \deptno, \annot [\vFive] \salary, \\
&\hspace{55pt} \annot [\edu] \isstudent,
\annot [\edu] \isteacher )^{\vFour \vee \vFive}\\
&\hspace{17pt} \ecourse ( \cno, \cname, \annot [\tFive] \deptno )^{\tFour \vee \tFive} \}^{\fModel_i}
\end{align*}
where the feature model $\fModel_i$ depends on whether the user wants to allow 
only one temporal feature for each schema column be enabled at a given time, $\fModel_1$,
or any number of them, $\fModel_2$. Hence, the feature model can capture exactly the
user need and how they want to encode variability within their DB.
\begin{align*}
\fModel_1 &= 
\paran {\neg \edu \wedge \paran {\paran {\vFour \wedge \neg \vFive} \vee {\vFive \wedge \neg \vFour}}}\\
&\hspace{3pt} \vee
\paran {\edu \wedge \paran {\paran {\vFour \wedge \neg \vFive} \vee {\vFive \wedge \neg \vFour}} \wedge
\paran {\paran {\tFour \wedge \neg \tFive} \vee \paran {\tFive \wedge \tFour}}}\\
\fModel_2 &= \paran {\neg \edu \wedge \paran {\vFour \vee \vFive}} \vee \paran {\edu \wedge \paran {\tFour \vee \tFive} \wedge \paran {\vFour \vee \vFive}}
\end{align*}
When the feature model is set to \ensuremath {\fModel_2}, 
the presented v-schema is compactly representing \ensuremath {20} schema variants, 
including four schema variants when \edu\ is disabled and 16 schema variants when \edu\
is enabled.
\end{example}

\textbf{Hierarchal structure of feature expressions in a v-schema:}
Since a v-schema is an annotated variational set, it
follows the properties 
of variational set stated in \secref{vlist-vset}
Thus, the feature model (the v-schema presence condition)
is enforced to all v-relations and the presence condition of a relation
is enforced to all its attributes. This means the real feature expression
of an attribute is the conjunction of its presence condition with
its relation presence condition and feature model.
Similarly,
the real feature expression of a relation is its presence condition
conjuncted with feature model.
In other words, the annotated attribute \optAtt\ of v-relation \vRel\ with 
$\dimMeta_r = \getPC \vRel$
defined in the v-schema \vSch\ with feature model \fModel\
is valid if: $\sat {\dimMeta \wedge \dimMeta_r \wedge \fModel }$.
For example, the \isstudent\ attribute described in \exref{vsch} 
is only valid if its presence condition is
satisfiable, i.e.,
\ensuremath {\sat {\edu \wedge (\vFour \vee \vFive) \wedge \fModel } = \t}.

\subsubsection{Configuring a V-Schema}
In essence, a v-schema is a systematic 
and compact representation of all schema variants of the 
underlying application of interest, e.g., SPL, that encodes 
the variability effectively inside the database schema by means of 
feature-related information (feature expressions).
As a result, it relieves the
need to define an intermediate schema and state mappings 
between it and source schemas, like the approach that 
data integration systems employ. 
However, the user can still obtain the specific pure relational schema for
a database variant by \emph{configuring} the v-schema with variant's configuration.
We formally define the configuration function for v-schemas and its elements in \figref{vsch-conf-sem}.
For example, consider the v-schema given in \exref{vsch}.
Configuring the variational attribute set of the \empacct\ relation for 
the variant \setDef {\vFive}, i.e., 
\ensuremath {\olSem [\setDef {\vFive}] {\empno, \hiredate, \titleatt, \deptno}},
is the relational attribute set of
\ensuremath {\setDef {\underline{ \empno, \hiredate, \titleatt, \deptno} }}.

\subsection{Variational Table}
\label{sec:vtab}

Variability can also exist in database content, i.e., tuples. To account 
for content variability, we tag tuples with 
presence conditions and in order to store it we move it into the attribute set
of a relation, resulting in \emph{variational tables (v-tables)}. 
To avoid overcrowding the database with feature 
expressions we only annotate tuples and not every single cell. 
This is feasible because we take advantage of the 
v-relation schema already assigned to a relation to ensure the value of a cell is 
valid, i.e., every cell has three presence conditions:
1) its attribute's presence 
condition, 2) its tuple's presence condition, and 3) the relation's presence condition.
The conjunction of these three
feature expressions must evaluate under a configuration for the cell value to
be valid. 

\begin{definition}[Variational table]
A \emph{variational table (v-table)}, \vTab, of relation 
$\pRelSch = \pRel \paran {\vi \pAtt k}$ has the v-relation
schema: 
$\vRelSch = \vRel \left( \optAtt [\dimMeta_1] [\vAtt_1], \hdots, \optAtt [\dimMeta_k] [\vAtt_k], \pc\right)^\dimMeta$.
A \emph{variational tuple (v-tuple)} is a mapping between a list of 
v-relation schema attributes, including the presence condition attribute, 
and their values, i.e., $t = \left( {\vi v k}, \dimMeta_t \right)$. Hence, 
a \emph{variational relation content}, \vRelCont, is a set of v-tuples 
$\synDef \vRelCont \vRelContSet \eqq \setDef {\vi \tuple n}$.
The \emph{v-table} \vTab\ is the pair of its relation schema \vRelSch\ and content \vRelCont.
Note that the value $v_i$ is valid and present iff 
$\sat {\dimMeta_i \wedge \dimMeta_\tuple \wedge \dimMeta \wedge \fModel} = \t$,
where as a reminder, 
$\dimMeta_i = \getPC {\getAtt v}$,
$\dimMeta_\tuple = \getPC \tuple$,
\dimMeta = \getPC \vRel,
and 
\fModel\ is the feature model.
\end{definition}



\begin{definition}[VDB instance]
\label{def:vdb-inst}
A \emph{variational database instance}, \vdbInst, of VDB \vDB, is an annotated
variational set of v-tables where the feature model of \vDB, i.e., \fModel, 
is the VDB instance's presence condition: 
$\synDef \vdbInst  \vdbInstSet \eqq \annot [\fModel] {\setDef {\vi \vRelCont l}} $.
\end{definition}

Similar to v-schema, a user can choose to configure a v-table or a VDB
for a specific variant,
elucidated in \secref{impl-var}.

\section{Variational Query}
\label{sec:vq}

To express variability in queries,
we utilize choice calculus to represent variability and incorporate it into a 
structured query language. We formally define 
\emph{variational relational algebra (VRA)}, \secref{vrel-alg},
as our algebraic query language.
A query written in VRA is called a \emph{variational query (v-query)},
when it is clear from the context we use query and v-query interchangeably. 
V-query typically conveys the same intent over several 
relational database variants, however, it is also capable of capturing different 
intents over database variants.
Consequently, the expressiveness of v-queries may cause them to be 
more complicated than relational queries, discussed in \secref{type-sys}. 
Hence, we introduce a 
\emph{type system} for VRA that statically checks if a 
v-query conforms to the underlying v-schema and encoded variability within the VDB.
Finally, we close out this section by providing a set of rules in \secref{var-min} 
for reducing a query's variability.

\subsection{Variational Relational Algebra}
\label{sec:vrel-alg}

Considering the variational nature of a VDB, to satisfy a user's information 
need when extracting information from a VDB, 
we need a query language that not only considers the structure of 
relational databases (such as SQL and relational algebra (RA)) but also 
accounts for the variability encoded in the VDB. We achieve this by:
1) picking relational algebra as our main query language and
2) using \emph{choices}~\cite{Walk13thesis, EW11tosem} 
and presence conditions to account for variability. 

A choice $\chc {x,y}$ consists of a feature expression, \dimMeta, and 
two alternatives $x$ and $y$. For a given configuration \config, 
the choice $\chc {x, y}$ can be replaced with $x$ if \dimMeta\
evaluates to \t\ under configuration \config, (i.e., \fSem \dimMeta = \t), 
or $y$ otherwise. 
In essence, choices allow v-queries (and its parts) 
to capture/encode variation in a structured and systematic manner. 

\begin{figure}
\textbf{Variational conditions:}
\begin{syntax}
\synDef \vCond \vCondSet &\eqq& \bTag \myOR \pAtt \bullet \cte \myOR \pAtt \bullet \pAtt \myOR 
						\neg \vCond \myOR \vCond \vee \vCond \\
                        &\myOR& \vCond \wedge \vCond
                         \myOR \chc { \vCond, \vCond}
\end{syntax}

\medskip
\textbf{Variational relational algebra syntax:}
\begin{syntax}
\synDef \vQ \qSet &\eqq& \vRel &\textit{Variational Relation}\\
								&\myOR& \vSel \vQ &\textit{Variational Selection}\\
                                &\myOR& \vPrj [\vAttList] \vQ &\textit{Variational Projection}\\
                                &\myOR&  \chc {\vQ,\vQ} &\textit{Variational Expression Choices}\\
			      &\myOR& \vQ \Join_\vCond \vQ &\textit{Variational Join}\\
			      &\myOR& \vQ \times \vQ &\textit{Variational Cartesian Product}\\
                                &\myOR& \vQ \circ \vQ &\textit{Variational Set Operation}\\
                                &\myOR& \empRel &\textit{Empty Relation}
\end{syntax}

\caption{Variational relational algebra definitions. $\circ$ denotes set operators: union and difference.}
\label{fig:v-alg-def}
\end{figure}

The syntax of variational relational algebra (VRA) 
is given in \figref{v-alg-def}.
The selection is adjusted to take a variational condition (i.e.,
a relational condition extended with choices).
For example, the query $\sigma_{\chc {\vAtt_1=\vAtt_2,\vAtt_1=\vAtt_3}} \vRel$
selects two groups of tuples: 1) tuples that satisfy 
conditions $\vAtt_1 = \vAtt_2$ and  \dimMeta\ 
and 
2) tuples that satisfy $\vAtt_1 = \vAtt_3$ and $\neg \dimMeta$.
A join of queries also takes a variational condition.
The projection takes a variational set of attributes, e.g.,
the query $\pi_{\vAtt_1, \optAtt [\dimMeta] [\vAtt_2]} \vRel$
projects $\vAtt_1$ unconditionally and $\vAtt_2$ only when \dimMeta\ is satisfied.
VRA allows for a choice of two v-queries. This demands an
\emph{empty} query since an alternative of a choice can very well inquire 
no information at all. The rest of VRA's operations are the same as RA.

%

\textbf{Expressiveness of VRA:}
VRA consists of all operators of RA in addition to some extra ones, i.e., VRA is a
conservative extension of RA. Hence, VRA is more expressive than RA due to 
its variational nature. This variational nature allows users to write interesting queries:
1) to express their exact information need they can use annotations or 
choices, \exref{vq-specific},
2) to express the same intent over several database variants they can 
use choices in queries or conditions, \exref{vq-same-intent-mult-vars},
and 
3) they can also use choices to express different intents over database variants.



\begin{example}
\label{eg:vq-specific}
Assume we have a VDB that includes features \vThree, \vFour, and \vFive\
with the \empbio\ table given in \tabref{mot}. The v-schema of such a VDB is:
\begin{align*}
\vSch_2 &=
\{\empbio (\empno, \sex, \birthdate,
\optAtt [\vFour] [\name], \optAtt [\vFive] [\fname],\\
& \hspace{50pt} \optAtt [\vFive] [\lname] )\}^{\vThree \vee \vFour \vee \vFive}.
\end{align*}
The user wants to get employees ID numbers and names for variants 
\vFour\ and \vFive.
The user needs to project the \name\ attribute 
for variant \vFour, the \fname\ and \lname\ attributes for variant \vFive,
and \empno\ attribute for both variants.
Hence, we have the v-query:\\
\centerline{
\ensuremath{
\vQ_1 = 
\pi_{\optAtt [\neg \vThree] [\empno], \optAtt [\vFour] [\name], \optAtt [\vFive] [\fname], \optAtt [\vFive] [\lname]  } \empbio.}}
The query can be simplified to \ensuremath{\vQ_2}
since feature expressions assigned to projected attributes in $\vQ_1$ conform
to attributes presence conditions in v-schema \ensuremath {\vSch_2} except  for
\optAtt [\neg \vThree] [\empno]:\\
\centerline{
\ensuremath{
\vQ_2 = \pi_{\optAtt [\neg \vThree] [\empno], \name, \fname, \lname} \empbio}}.
Hence, the user does not have to repeat the variability encoded in the VDB in their
query again. We discuss this more in \secref{type-sys} and \exref{type-sys}. 
\end{example}

\begin{example}
\label{eg:vq-same-intent-mult-vars}
If the user desires to get all employee 
names, i.e., for all temporal variants \vOne -- \vFive, of our motivating example
they can query:\\
\ensuremath{
\vQ_3 = \chc [\vOne] {\pi_{\name} \engemp \cup \pi_{\name} \othemp, 
\chc [\vTwo \vee \vThree ] {\pi_{\name} \empacct,
\chc [\vFour \vee \vFive] {\pi_{\name, \fname, \lname} \empbio, \emp}
}
} }. 
The v-schema enforces that if the first three variants are not enabled one of 
the last two are. Hence, we can simplify $\vQ_3$ to:\\
$\vQ_4 = \chc [\vOne] {\pi_{\name} \engemp \cup \pi_{\name} \othemp, 
\chc [\vTwo \vee \vThree ] {\pi_{\name} \empacct,
\pi_{\name, \fname, \lname} \empbio
}
} $.
Expressing the same intent over several instances by a single query relieves the
DBA from rewriting a query multiple time based on the schema of the underlying variant.
\end{example}


\textbf{Running a v-query on a VDB results in a v-table:}
A v-query systematically represents a set of query variants written for their
corresponding database variants. Hence, intuitively the user expects to 
get such variability in their result as well. A v-table captures the 
result of a v-query over a VDB. 
Its relation schema determines the structure of the table and the condition
under which they are valid and present (i.e., when their presence condition is satisfiable).
It captures the variability enforced
by the query in both the relation schema of the v-table and its content.
Additionally, the presence condition of tuples track the variants where they are
present, i.e., they do not lose the meta-information of what variant a tuple
belongs to (data provenance). 
Note that the presence condition of tuples, attributes, and the return relation
is restricted due to the variability enforced by the query. 
For example, assume the tuple 
$(1,2,\dimMeta_1)$ belongs to v-relation $\vRel (\vAtt_1, \vAtt_2)$.
The query $\pi_{\optAtt [\dimMeta] [\vAtt_1]} \vRel$ returns a v-table that
consists of variational attribute set $\optAtt [\dimMeta] [\vAtt_1]$ and it 
includes the tuple $(1, \dimMeta_1 \wedge \dimMeta)$. The 
$\dimMeta_1$ part of tuple's presence condition states the origin of the tuple
while the \dimMeta\ part describes the restriction of the query applied to the tuple.

\textbf{Renaming of queries and attributes:}
In our implementation, VRA also allows for renaming of attributes and queries 
similar to relational algebra. However, to avoid unnecessary complexity we 
omit it from the formal definition in \figref{v-alg-def}.

\subsubsection{VRA Semantics}
\label{sec:vra-sem}
To define VRA semantics we can map 
a v-query to a pure relational query to re-use RA's semantics.
However, to avoid losing information, i.e., the variability encoded 
in the v-query which does not exist in RA queries, 
we need to determine the variant under which such a
mapping is valid. Hence, we introduce two semantic functions that 
relate a v-query to a relational query.

\textbf{Configuring a v-query:} 
maps a v-query under a 
given configuration to a relational query, denoted by \eeSem . 
and defined in \figref{v-alg-conf-sem}. Hence, configuring a v-query
for all valid configurations, accessible from VDB's feature model,
provides the complete meaning of a v-query in terms of RA semantics.

\begin{example}
\label{eg:conf-vq}
Assume the underlying VDB has \t\ feature model and the v-relation
\ensuremath{
\vRel \left( \optAtt [\fOne] [\vAtt_1], \vAtt_2, \vAtt_3 \right)^{\fOne \vee \fTwo}
} and only two features \fOne\ and \fTwo.
The v-query 
\ensuremath{
\vQ_5 = \vPrj [{\vAtt_1, \optAtt [\fOne \wedge \fTwo] [\vAtt_2], \optAtt [\fTwo] [\vAtt_3]}] \vRel
}
is configured to the following relational queries:
\ensuremath{\eeSem [\setDef \fOne] {\vQ_5} = \eeSem [\setDef \ ] {\vQ_5} = \pi_{\pAtt_1} \pRel},
\ensuremath{\eeSem [\setDef \fTwo] {\vQ_5} =
 \pi_{\pAtt_1, \pAtt_3} \pRel},
\ensuremath{\eeSem [\setDef {\fOne, \fTwo}] {\vQ_5} = \pi_{\pAtt_1, \pAtt_2, \pAtt_3} \pRel}.
\end{example}


\begin{figure}
\begin{alignat*}{1}
\eeSem [] . &: \qSet \to \confSet \to \pQSet\\
\eeSem \vRel &= \orSem \vRel = \pRel\\
\eeSem {\vSel \vQ}  &= \vSel [\ecSem \vCond] {\eeSem \vQ}\\
\eeSem {\vPrj [\vAttList] \vQ} &= \vPrj [\olSem \vAttList] {\eeSem \vQ}\\
\eeSem {{\vQ_1} \times {\vQ_2}} &= \eeSem {\vQ_1} \times \eeSem {\vQ_2}\\
\eeSem {{\vQ_1} \Join_\vCond {\vQ_2}} &= \eeSem {\vQ_1} \Join_{\ecSem \vCond} \eeSem {\vQ_2}\\
\eeSem {\chc {\vQ_1, \vQ_2}} &= 
	\begin{cases}
		\eeSem {\vQ_1}, \text{ if } \fSem \dimMeta = \t\\
		\eeSem {\vQ_2}, \text{ otherwise}
	\end{cases}\\
\eeSem {{\vQ_1} \circ {\vQ_2}} &= \eeSem {\vQ_1} \circ \eeSem {\vQ_2}
\end{alignat*}
\caption{Configuration of VRA which assumes that the given v-query
is well-typed. \orSem ., \ecSem ., and \olSem . are
configuration of v-relation, v-condition, and variational attribute
set, respectively, defined in \figref{vsch-conf-sem}, 
\figref{vcond-conf-sem}, \figref{vsch-conf-sem}.}
\label{fig:v-alg-conf-sem}
\end{figure}

\textbf{Grouping a v-query :} 
maps a v-query to a set of
relational queries annotated with feature expressions, denoted by \qGroup .
and defined in \figref{vq-group}. The presence condition of relational queries 
indicate the group of configurations where the mapping holds. In essence, 
grouping of v-query \vQ\ groups together all configurations with the same relational
query produced from configuring \vQ. 
Hence, the generated set
of relational queries from grouping a v-query contains distinct (unique) queries.
For example, consider the query \ensuremath {\vQ_5} in \exref{conf-vq}.
Grouping \ensuremath{\vQ_5} results in the set:
\ensuremath{
\setDef{
\left( \pi_{\pAtt_1, \pAtt_2, \pAtt_3} \pRel \right)^{\fOne \wedge \fTwo},
\left(\pi_{\pAtt_1, \pAtt_3} \pRel \right)^{\neg \fOne \wedge \fTwo},
\left(  \pi_{\pAtt_1} \pRel \right)^{( \fOne \wedge \neg \fTwo) \vee (\neg \fOne \wedge \neg \fTwo)}
}
}.

\begin{figure}
\begin{alignat*}{1}
\qGroup . &: \qSet \to \ddot \pQSet\\
\qGroup \vRel &= \annot [\t] \pRel\\
\qGroup {\vSel \vQ}  &=  
\setDef {\annot [\dimMeta \wedge \dimMeta_\vCond] {\left(\sigma_{\pCond} \pQ\right)} \myOR
\annot \pQ \in \qGroup \vQ, \annot [\dimMeta_\vCond] \pCond \in \cGroup}
\\
\qGroup {\vPrj [\vAttList] \vQ} &= 
\setDef {\annot [\dimMeta \wedge \dimMeta_\vAttList] {\left(\pi_{\pAttList} \pQ \right)} \myOR
\annot \pQ \in \qGroup \vQ, \annot [\dimMeta_\vAttList] \pAttList \in \aGroup}
\\
\qGroup {{\vQ_1} \times {\vQ_2}} &= 
\setDef {\annot [\dimMeta_1 \wedge \dimMeta_2] {\left(\pQ_1 \times \pQ_2\right)} \myOR
\annot [\dimMeta_1] \pQ_1 \in \qGroup {\vQ_1}, \annot [\dimMeta_2] \pQ_2 \in \qGroup {\vQ_2} }
\\
\qGroup {{\vQ_1} \Join_\vCond {\vQ_2}} &= 
\setDef {\annot [\dimMeta_1 \wedge \dimMeta_2 \wedge \dimMeta_\vCond] {\left(\pQ_1 \Join_{\pCond} \pQ_2 \right)} \myOR 
\annot [\dimMeta_1] \pQ_1 \in \qGroup {\vQ_1}, \annot [\dimMeta_2] \pQ_2 \in \qGroup {\vQ_2}\\
& \hspace{104pt}
,\annot [\dimMeta_\vCond] \pCond \in \cGroup  }
\\
\qGroup {\chc {\vQ_1, \vQ_2}} &= 
\setDef {\annot [\dimMeta \wedge \dimMeta_1] \pQ_1 \myOR  \annot [\dimMeta_1] \pQ_1 \in \qGroup {\vQ_1} }
\cup 
\setDef {\annot [\neg \dimMeta \wedge \dimMeta_2] \pQ_2 \myOR  \annot [\dimMeta_2] \pQ_2 \in \qGroup {\vQ_2}}  \\
\qGroup {{\vQ_1} \circ {\vQ_2}} &= 
\setDef {\annot [\dimMeta_1 \wedge \dimMeta_2] {\left(\pQ_1 \circ \pQ_2\right)} \myOR
\annot [\dimMeta_1] \pQ_1 \in \qGroup {\vQ_1}, \annot [\dimMeta_2] \pQ_2 \in \qGroup {\vQ_2} }
\end{alignat*}
\caption{Grouping of v-queries. \cGroup\ and \aGroup\ indicate the grouping of v-conditions and
variational attribute lists, respectively. They both follow the definition provided in 
\secref{group}.
}
\label{fig:vq-group}
\end{figure}

\subsection{Well-Typed (Valid) V-Query}
\label{sec:type-sys}

To prevent running v-queries that have some sort of error
we implement a \emph{static type system} for VRA. The 
type system ensures queries are \emph{well-typed}, i.e., they conform
to the underlying v-schema both w.r.t. traditional structure of 
the database and the variability encoded in the database. 
For example, while projecting an annotated attribute \optAtt\ from a 
v-relation \vRel\ not only the attribute must belong to the v-relation, i.e., 
$\vRel \annot [\dimMeta_\vRel] {\paran {\optAtt [\dimMeta_1], \vAttList}}$, but 
the feature expression $\dimMeta \wedge \dimMeta_\vRel \wedge \dimMeta_1$ 
must also be satisfiable, i.e., 
the attribute \vAtt\ must be present in the relation \vRel\
under the condition imposed by the query and the v-schema. 
The type system is designed s.t. it relieves the user from necessarily incorporating
the v-schema variability into their queries as long as the v-queries variability
does not violate the v-schema, i.e., the variability encoded in v-queries can
be more restrictive or more loose than v-schema variability without violating them.
This is illustrated in \exref{vq-specific} and is explored more in \exref{type-sys}.
%

\begin{figure}

\textbf{V-queries typing rules:}

  \begin{mathpar}
  \small
  
    \inferrule[\relationE]
  	{\vRel (\vType)^{\VVal \dimMeta} \in \vSch \\
	\sat{\vctx \wedge \VVal \dimMeta} }
     {\envWithSchema{\envInContext [\vctx \wedge \VVal \dimMeta] {\vType}}}
  
  \inferrule[\prjE]
  	{\envPrime \\
    	\subsume {\annot \vType}  {\annot [\VVal \vctx] {\VVal \vType}}}
    {\env{\vPrj[\vType] \vQ} {\envInContext [\VVal \vctx] {\left(\annot {\vType} \cap {\VVal \vType} \right)}}}

  \inferrule[\selE]
  	{\env \vQ {\envInContext [\VVal \vctx] \vType} \\
    	\envCondAnnot \vCond}
    {\env{\vSel \vQ}{\envInContext [\VVal \vctx] \vType}}
    
  \inferrule[\choiceE]
  	{\envOne[\vctx \wedge \VVal \dimMeta] \\
    	\envTwo[\vctx \wedge \neg \VVal \dimMeta]}
    {\env{\chc[\VVal \dimMeta]{\vQ_1, \vQ_2}}{
     \envInContext [\vctx_1 \vee \vctx_2] {\left({\envInContext [\vctx_1] \vType_1} \cup
    							{\envInContext [\vctx_2] \vType_2}\right)}}}
    
  \inferrule[\productE]
  	{\envOne \\
    	\envTwo\\
	\annot [\vctx_1] \vType_1 \cap \annot [\vctx_2] \vType_2 = \{\}}
    {\env{\vQ_1 \times \vQ_2}{\envInContext [\vctx_1 \wedge \vctx_2] {\left(\vType_1 \cup \vType_2\right)}}}

  \inferrule[\setopE]
  	{\envOne \\
    	\envTwo \\
	\envEval {\annot [\vctx_1] \vType_1} {\annot [\vctx_2] \vType_2}}
    {\env{\vQ_1 \circ \vQ_2} {\envInContext [\vctx_1] \vType_1} }

  \end{mathpar}
  
\medskip
\textbf{V-condition typing rules (b: boolean tag, \pAtt: plain attribute, k: constant value):}
  \begin{mathpar}
  \small    

  \inferrule[\conjC]
  	{\envCond \vCond_1\\
    	\envCond \vCond_2}
    {\envCond{\vCond_1 \wedge \vCond_2}}
    
  \inferrule[\disjC]
  	{\envCond \vCond_1\\
    	\envCond \vCond_2}
    {\envCond{\vCond_1 \vee \vCond_2}}

  \inferrule[\choiceC]
    	{\envCond[\vctx \wedge \VVal \dimMeta, \vType]{\vCond_1} \\
        \envCond[\vctx \wedge \neg \VVal \dimMeta, \vType]{\vCond_2}}
    {\envCond{\chc[\VVal \dimMeta]{\vCond_1, \vCond_2}}}

  \inferrule[\notC]
  	{\envCond \vCond}
    {\envCond \neg \vCond}
        

  \inferrule[\attValC]
  	{
    	\optAtt [\VVal \dimMeta] \in \vType \\
	\taut{{\VVal \dimMeta} \imply \vctx} \\
        \cte \in \dom \vAtt}
    {\envCond{\op \pAtt \cte}}
    
  \inferrule[\boolC]
  	{}
    {\envCond \bTag}

  \inferrule[\attAttC]
  	{
    	\optAtt [\dimMeta_1] [\vAtt_1]\in \vType \\
         {\optAtt [\dimMeta_2] [\vAtt_2]} \in \vType \\
         \taut{\dimMeta_1 \imply \vctx} \\
         \taut{\dimMeta_2 \imply \vctx} \\
        \type[\vAtt_1] = \type[ \vAtt_2]}
    {\envCond{\op{\pAtt_1}{\pAtt_2}}}
    
  \end{mathpar}

\caption{VRA and v-condition typing relation. The typing rule of a join query is the combination
of rules \selE\ and \productE.}
\label{fig:vq-stat-sem}
\end{figure}

\figref{vq-stat-sem} defines VRA's \emph {typing relation}
as a set of inference rules assigning \emph{types} to queries. 
The type of a query is an annotated variational set of attributes.
The set of attributes are produced by running the query and
their presence conditions 
determine for what variants 
they are valid which vary from the original presence conditions of 
attributes and tables encoded in the VDB due to variability constraints
imposed by the query.
The presence condition of the entire set determines the condition under
which the entire table (i.e., attributes and tuples) are valid. 
Note that it is essential to consider the type of a query an \emph{annotate}
variational set to account for the presence condition of the entire table. If we
consider the type of a query a variational attribute set we lose information (i.e.,
the condition under which tuples are valid).
To capture the variability encoded in a query,
we keep and refine a feature expression, called a \emph{variation context}
and denoted by \vctx, while type checking and running v-queries. 
The variation context is first initiated by the feature model of the 
v-schema. 


%
VRA's typing relation, as defined in \figref{vq-stat-sem}, 
has the judgement form \env \vQ {\envInContext [\VVal \vctx] \vType}. 
This represents
that in \vctxTxt\ \vctx\ within v-schema \vSch, 
\vqTxt\ \vQ\ has type \envInContext [\VVal \vctx] \vType. 
If a query does not have a type, it is \emph{ill-typed}.
The rule \relationE\ states that, in \vctxTxt\ \vctx\ with
underlying \vschTxt\ \vSch, assuming that
1) \vSch\ contains
the relation \vRel\ with presence condition $\VVal \dimMeta$
and variational attribute set \vType\ 
and
2) there exists a valid variant in the intersection of variation context \vctx\
and \vRel's presence condition \ensuremath{\VVal \dimMeta}, i.e., 
\ensuremath {\sat {\vctx \wedge \VVal \dimMeta}},
then query \vRel\ has type \vType\ annotated with \ensuremath { \vctx \wedge \VVal \dimMeta}.

%
The rule \prjE\ states that, in \vctxTxt\ \vctx\ within v-schema \vSch, assuming 
that the subquery \vQ\ has type $\envInContext [\VVal \vctx] {\VVal \vType}$,
\vqTxt\ $\pi_\vType \vQ$
has type \ensuremath {\envInContext [\VVal \vctx] {\left( \annot \vType \cap \VVal \vType\right)}}, 
if \ensuremath {\envInContext [\VVal \vctx] {\VVal \vType}} subsumes the variational 
attribute set \vType\ constrained (annotated) with variational context \vctx, i.e., \annot \vType.
The subsumption, defined in \defref{vset-subsumption},
ensures that the subquery \vQ\ includes all attributes in 
the projected attribute set and attributes' presence conditions do not 
contradict each other. Returning the intersection of types, defined in 
\defref{vset-intersect}, filters both 
attributes and their presence conditions.

\begin{example}
\label{eg:type-sys}
Consider \exref{vq-specific}. 
\ensuremath{\vQ_1} and \ensuremath{\vQ_2}
represent the same information need and have the same type:
1) The subquery \empbio\ 
has the type:
\ensuremath{
\vAttList =  \{\empno, \sex, \birthdate,}
\ensuremath{ 
\optAtt [\vFour] [\name], \optAtt [\vFive] [\fname], \optAtt [\vFive] [\lname]\}^{\vThree \vee \vFour \vee \vFive }},
in variational context \ensuremath{\vThree \vee \vFour \vee \vFive} within schema \ensuremath{\vSch_2}.
2) Both projected attribute lists of \ensuremath{\vQ_1} and \ensuremath{\vQ_2}, i.e.,
\ensuremath{
{\setDef {\optAtt [\neg \vThree] [\empno],
\optAtt [\vFour] [\name], \optAtt [\vFive] [\fname], \optAtt [\vFive] [\lname]}}}
and
\ensuremath{
{\setDef {\optAtt [\neg \vThree] [\empno],
\optAtt [\vFour] [\name], \optAtt [\vFive] [\fname], \optAtt [\vFive] [\lname]}}}, respectively,
are subsumed 
by \vAttList\ because \vAttList\ contains all elements of projected lists and the conjunction
of presence conditions from \vAttList\ and the projected list are satisfiable. 
3) The intersection of both 
projected attribute lists of \ensuremath{\vQ_1} and \ensuremath{\vQ_2} 
with \vAttList\ is:\\
\ensuremath{
{\setDef {\optAtt [\neg \vThree] [\empno],
\optAtt [\vFour] [\name], \optAtt [\vFive] [\fname], \optAtt [\vFive] [\lname]}}^{\vThree \vee \vFour \vee \vFive}}.
\end{example}

%
The rule \selE\ states that, in \vctxTxt\ \vctx\ within v-schema \vSch, assuming 
that the subquery \vQ\ has type {\envInContext [\VVal \vctx] \vType}, 
the v-query $\sigma_{\vCond} \vQ$
has type {\envInContext [\VVal \vctx] \vType},
if the \vCondTxt\ \vCond\ is well-formed w.r.t.
 \vctxTxt\ \vctx\ and \tenvTxt\ {\envInContext [\VVal \vctx] \vType}, 
denoted by v-condition's typing relation 
\envCondAnnot \vCond.
Note that in variational condition typing rules, 
the presence condition of the query type is applied to the 
variational attribute set, thus, they have the form \envCond \vCond\ 
instead of \envCondAnnot \vCond. 
The rules state that attributes used in a
\vCondTxt\ must be valid in \vType\ and 
attribute's \presCondTxt\ \ensuremath {\VVal \dimMeta} 
in type \vType\ must be more specific than the \vctxTxt\ \vctx,
denoted by \ensuremath{\taut {\VVal \dimMeta \to \vctx}}, 
since \vType\ is the exact type and specification of the subquery within
a selection query which is at least as specific as the \vctxTxt\ under which
the selection query is written. 
They also
check the constraints of traditional relational databases, such as the type of two 
compared attributes must be the same.

%
The rule \choiceE\ states that, in \vctxTxt\ \vctx\ within v-schema \vSch, the type of 
a choice of two subqueries is the \emph{union of types}, defined in 
\defref{vset-union}, of its subqueries annotated with the disjunction of their presence
conditions.
A choice query is well-typed iff both of 
its subqueries $\vQ_1$ and $\vQ_2$ are well-typed.
Note that we do not simplify 
\ensuremath{
\annot [\vctx_1 \vee \vctx_2] {\left(\envInContext [\vctx_1] \vType_1 \cup \envInContext [\vctx_2] \vType_2\right)}
}
to 
\ensuremath{
\envInContext [\vctx_1] \vType_1 \cup \envInContext [\vctx_2] \vType_2
}
because we need to know the presence condition of the entire type, i.e., \ensuremath{\vctx_1 \vee \vctx_2},
to know the condition under which tuples are valid\footnote{The simplification holds because
\ensuremath{
\annot [\vctx_1 \vee \vctx_2] {\left(\envInContext [\vctx_1] \vType_1 \cup \envInContext [\vctx_2] \vType_2\right)}
\equiv
\envInContext [\vctx_1\wedge (\vctx_1 \vee \vctx_2)] \vType_1 \cup \envInContext [\vctx_2\wedge (\vctx_1 \vee \vctx_2)] \vType_2
\equiv
\envInContext [\vctx_1] \vType_1 \cup \envInContext [\vctx_2] \vType_2
}
}.

%
The rule \productE\ states that the type of a product query in \vctxTxt\
\vctx\ is the union of the type of its subqueries annotated with the 
conjunction of their presence conditions, assuming that 
they are disjoint. Note that 
\ensuremath{
\annot [\vctx_1 \wedge \vctx_2] {\left(\envInContext [\vctx_1] \vType_1 \cup \envInContext [\vctx_2] \vType_2\right)}
\equiv 
\envInContext [\vctx_1 \wedge \vctx_2] \vType_1 \cup \envInContext [\vctx_1 \wedge \vctx_2] \vType_2
\equiv
\annot [\vctx_1 \wedge \vctx_2] {\left(\vType_1 \cup \vType_2\right)}
}.

%
The rule \setopE\ denotes the typing rule for set operation queries such as 
union and difference. It states that, if the subqueries $\vQ_1$ and $\vQ_2$
have types $\envInContext [\vctx_1] \vType_1$ and 
$\envInContext [\vctx_2] \vType_2$, respectively, in \vctxTxt\ \vctx,
then the \vqTxt\ of their set operation has type $\envInContext [\vctx_1] \vType_1$, iff 
$\envInContext [\vctx_1] \vType_1$ and $\envInContext [\vctx_2] \vType_2$ are \emph{equivalent}.
The \emph{type equivalence} is v-set equivalence, defined in \defref{vset-eq},
for v-sets of attributes.

\textbf{Variation-preserving property:}
VRA's 
type system is \emph{variation-preserving}, 
i.e., the type system takes
a v-query and produces correspondingly variational attribute set (the
type of a v-query) w.r.t. the underlying v-schema.
\begin{wrapfigure}{r}{0.12\textwidth}
\begin{center}
\begin{tikzcd}[column sep=2.3em]
  \vQ   \rar{\mathit{type}}  \dar[swap,dashed]{\sem[c]{\cdot}}
& {\annot \vType}  \dar[dashed]{\sem[c]{\cdot}} \\
  \pQ \rar{\underline{\mathit{type}}}
& \pAttList 
\end{tikzcd}
\end{center}
\end{wrapfigure}
Assume \ensuremath{\mathit{type}(\vQ)} indicates the type of v-query \vQ\
while \ensuremath{\underline{\mathit{type}}(\pQ)} indicates the type of relational query \pQ.
Simply put, 
the relational type of the configured v-query \vQ\ with configuration \config, 
i.e., \ensuremath{\olSem {\mathit{type} (q)}},
must be the same as the configured variational type 
of the v-query \vQ\ with configuration \config, 
i.e., \ensuremath{\underline {\mathit{type}} (\eeSem {\vQ})}.
As shown in the diagram, taking either path of 1) configuring \vQ\ first and 
then getting the relational type of it and 
2) getting the variational type of \vQ\ first and then configuring it results
in the same set of attributes. 
%
Variation-preserving property of VRA's type system and RA's type safety 
implies that VRA's type system is also type safe.

\subsection{Variation Minimization}
\label{sec:var-min}


VRA is flexible since an information need can be represented via multiple
v-queries as demonstrated in \exref{vq-specific} and \exref{vq-same-intent-mult-vars}.
It allows users to incorporate their personal taste and task requirements
into v-queries they write by 
having different levels of variation. For example, consider the query
\ensuremath {
\vQ_1 =  \pi_{\optAtt [\neg \vThree] [\empno], \optAtt [\vFour] [\name], \optAtt [\vFive] [\fname], \optAtt [\vFive] [\lname]  } \empbio} 
from \exref{vq-specific}. To be explicit about the exact query that will be run for 
each variant and knowing that 
\ensuremath{
\getPC \empbio = \vThree \vee \vFour \vee \vFive
},
the user can lift up the variation and rewrite the query as\\
\ensuremath{
\VVal \vQ_1 = \chc [\vFour] {\pi_{\empno, \name} \empbio, 
\chc [\vFive] {\pi_{\empno, \fname, \lname} \empbio, \emp}} 
}.
While \ensuremath{\vQ_1} contains less redundancy \ensuremath{\VVal \vQ_1}
is more comprehensible. 
\emph{Supporting multiple levels of variation 
creates a tension between reducing redundancy and maintaining comprehensibility.}

We define \emph{variation minimization} rules and include 
interesting ones in \secref{var-min}.
Pushing in variation (feature expressions) into a query, i.e., applying rules left-to-right, 
reduces redundancy and improves performance
while lifting them up, i.e., applying rules right-to-left, 
makes a query more understandable. 
When applied left-to-right, the rules are terminating since the scope of variation 
always gets smaller.

\begin{figure}
\textbf{Choice Distributive Rules:}
\begin{alignat*}{1}
\small
\chc {\pi_{\vAttList_1} \vQ_1, \pi_{\vAttList_2} \vQ_2}
&\equiv
\pi_{\annot \vAttList_1, \annot [\neg \dimMeta] \vAttList_2} \chc {\vQ_1, \vQ_2}\\
\chc {\sigma_{\vCond_1} \vQ_1, \sigma_{\vCond_2} \vQ_2} 
&\equiv
\sigma_{\chc {\vCond_1, \vCond_2}} \chc {\vQ_1, \vQ_2}\\
\chc {\vQ_1 \times \vQ_2, \vQ_3 \times \vQ_4}
&\equiv
\chc {\vQ_1, \vQ_3} \times \chc {\vQ_2, \vQ_4}\\
\chc {\vQ_1 \Join_{\vCond_1} \vQ_2, \vQ_3 \Join_{\vCond_2} \vQ_4}
&\equiv
\chc {\vQ_1, \vQ_3} \Join_{\chc {\vCond_1, \vCond_2}} \chc {\vQ_2, \vQ_4}\\
\chc {\vQ_1 \circ \vQ_2, \vQ_3 \circ \vQ_4}
&\equiv
\chc {\vQ_1, \vQ_3} \circ \chc {\vQ_2, \vQ_4}
\end{alignat*}

\medskip
\textbf{CC and RA Optimization Rules Combined:}
\begin{alignat*}{1}
\small
\chc {\sigma_{\vCond_1 \wedge \vCond_2} \vQ_1, \sigma_{\vCond_1 \wedge \vCond_3} \vQ_2}
&\equiv
\sigma_{\vCond_1 \wedge \chc {\vCond_2, \vCond_3}} \chc {\vQ_1, \vQ_2}\\
\sigma_{\vCond_1} \chc {\sigma_{\vCond_2} \vQ_1, \sigma_{\vCond_3} \vQ_2}
&\equiv
\sigma_{\vCond_1 \wedge \chc {\vCond_2, \vCond_3}} \chc {\vQ_1, \vQ_2}\\
\chc {\vQ_1 \Join_{\vCond_1 \wedge \vCond_2} \vQ_2, \vQ_3 \Join_{\vCond_1 \wedge \vCond_3} \vQ_4}
&\equiv
\sigma_{\chc {\vCond_2, \vCond_3}} \left( \chc {\vQ_1, \vQ_3} \Join_{\vCond_1} \chc {\vQ_2, \vQ_4} \right)
\end{alignat*}

\caption{Some of variation minimization rules.}
\label{fig:var-min}
\end{figure}

\section{VDBMS Implementation}
\label{sec:impl}


To interact with VDBs using v-queries, we implement 
\emph{Variational Database Management System (VDBMS)}.
VDBMS is extensibly implemented in Haskell s.t. it can sit on 
top of any DBMS that the user desires and the VDB is stored at. 
To acquire an extensible system we implement a shared interface
for connecting to and inquiring information from a DBMS and
instantiate it for different database engines such as PostgreSQL and
MySQL.

\begin{figure}
\includegraphics[width = \linewidth] {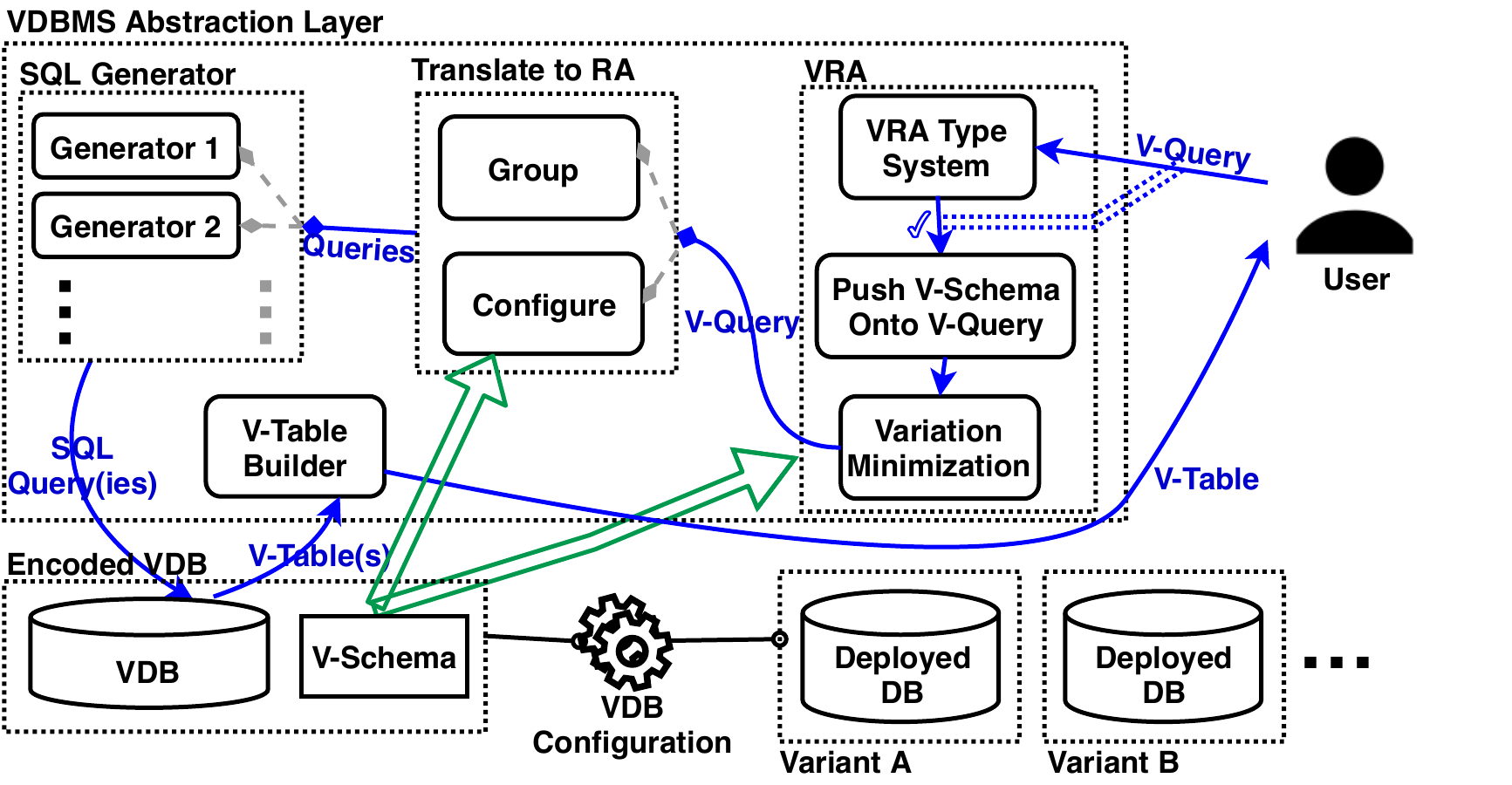}
\caption{VDBMS architecture. 
The dotted double-line from v-query to pushing v-schema module
indicates the dependency of passing the v-query to this module
only if it is well-typed. 
The dashed gray arrows with diamond heads demonstrate
an option for the flow of inputs. 
The blue filled arrows track the data flow, the green hollow arrows 
indicate an input to a module.}
\label{fig:arch}
\end{figure}

\figref{arch} shows the architecture of VDBMS and its modules.
For now, we assume a VDB and its v-schema are generated by an 
expert and stored in a DBMS. Later, we discuss generating VDBs in 
\secref{gen-vdb}. A VDB can be \emph{configured} to its pure relational 
database variants, if desired by a user, by providing the configuration
of the desired variant, elucidated in \appref{conf-vdb}.
For example, a SPL developer configures a VDB to produce 
software and its database for a client.

Given a VDB and its v-schema, a user inputs v-query \vQ\ to VDBMS.
First, \vQ\ is type-checked by the VRA type system introduced in 
\secref{type-sys}. If it is well-typed, 
to ensure variation-preserving property throughout the flow of v-query 
in the system, the \emph{v-schema is pushed onto the v-query} by
conjuncting attributes annotations with their presence conditions 
from the v-schema.
\exref{var-pres} elucidates the role of the push v-schema onto
the v-query.
It is then passed to the variation minimization module, introduced in 
\secref{var-min}, to minimize the variation of \vQ\ and apply
relational algebra optimization rules. 
If it is ill-typed, the user gets errors explaining what part of the 
query did not conform to the v-schema.
To generate runnable queries w.r.t. the underlying DBMS,
we pass the minimized query \ensuremath {\VVal \vQ} to 
the \emph{translate to RA} module that provides two approaches 
(\emph{grouping} and \emph{configuring}, explained in 
\secref{config} and \secref{group},
respectively) to generate RA queries. The generated 
queries are then sent to the \emph{SQL generator} module which generates
SQL queries in various ways from the relational algebra queries, explained
in \secref{sql-gen}.

\begin{example}
\label{eg:var-pres}
Consider the v-query \ensuremath{\vQ_5} given in \exref{conf-vq}. It is well-typed
because it has the type
\ensuremath{\vAttList =
\setDef {\optAtt [\fOne] [\vAtt_1], 
\optAtt [\fOne \wedge \fTwo] [\vAtt_2], 
\optAtt [\fTwo] [\vAtt_3]}
}.
Configuring \vAttList\ for the variant that both \fOne\ and \fTwo\ are disabled
results in an empty attribute set. However, the type of its configured query
for this variant, i.e., \ensuremath{\eeSem [\setDef \ ] {\vQ_5}}, is the 
attribute set \ensuremath{\setDef {\pAtt_1}}. This violates the
variation-preserving property. A similar problem happens for the variant of
\setDef {\fTwo}, i.e., \ensuremath{
\underline{\mathit{type}} \left( \eeSem [\setDef \fTwo] {\vQ_5} \right) = 
\setDef{\pAtt_1, \pAtt_3} \not = \setDef{\pAtt_3} = \olSem [\setDef \fTwo] {\mathit{type} \left( \vQ_5 \right)}
}.
We can restrict VRA's type system to enforce users to incorporate the
v-schema into their queries, e.g., \ensuremath{\vQ_5} becomes
\ensuremath{\VVal \vQ_5 = 
\vPrj [{\optAtt [\fOne] [\vAtt_1], 
\optAtt [\fOne \wedge \fTwo] [\vAtt_2], 
\optAtt [\fTwo] [\vAtt_3]}] \vRel
}. However, one of the purposes of our type system is to relieve the users 
from having to encode the VDB's variability into their queries.
To avoid this violation without requiring users to repeat VDB's variability in their queries,
after type checking a query we push the v-schema onto the v-query,
e.g., doing so for \ensuremath{\vQ_5} results in \ensuremath{\VVal \vQ_5}.
\end{example}

Having generated SQL queries, we now run them over the underlying 
VDB (stored in a DBMS desired by the user). The result could be either 
a v-table or a list of v-tables, depending on the approach chosen in 
the translator to RA and SQL generator modules. The v-table(s) is passed
to the \emph{v-table builder}
to create one v-table that filters out 
duplicate or invalid tuples and shrinks presence conditions and 
eventually, returns the final v-table to the user.

\subsection{Implementing A Variational Entity}
\label{sec:impl-var}

Recognizing that all entities in VDBMS including tables, schemas, VDBs, 
and queries are variational,
we implement the \emph{variational} shared interface to consists of all methods
that variational objects share. As illustrated through \secref{vdb} and \secref{vra-sem}, 
a variational object has two main methods:
1) \emph{configure}: given a variational object and a configuration returns the
non-variational object variant, i.e., configure is a function 
$\elemSem . : \mathbf{O} \to \confSet \to \underline{\mathbf{O}}$
and 
2) \emph{group}: given a variational object returns groups of annotated 
non-variational objects with a feature expression where the feature expression
describes the group of configurations where the object is present, i.e., 
group is a function 
\ensuremath{\elemGroup [.] : \mathbf{O} \to \ddot {\underline {\mathbf{O}}}}. 
Remember,
annotated objects are still variational, however, the group method removes 
any kinds of variability in a variational object and re-introduces variability only
by annotating non-variational objects.
Examples of configuring and grouping v-queries are given in \secref{vra-sem}.

\subsubsection{Configuring A Variational Entity}
\label{sec:config}

Conceptually, a variational entity represents many different
non-variational entities that can be obtained by configuration,
as shown already for v-schema, VDB, and v-query.
Hence, we define the \emph{configure} method
for all instances of the variational interface. 


\subsubsection{Grouping A Variational Entity}
\label{sec:group}

 
In essence, while configuring a variational entity may generate duplicate non-variational
entities, grouping a variational entity gathers the unique non-variational entities
and annotates them with a feature expression that represent the group of configurations where 
the non-variational entity is the result of configuring the variational entity with that configuration,
i.e., it moves the variation to the top level of a variational entity.
Hence, a variational entity \elem\ can be grouped into multiple non-variational entities annotated
with a feature expression \dimMeta, where \dimMeta\  represents the set (group) of configurations \confSet\
s.t. $\forall \config \in \confSet: \fSem \dimMeta = \t$. Thus, we define the \emph{group} 
method for the variational interface and formulate it for the variational entity \elem\ as:
$\elemGroup = \setDef {\annot \pElem \myOR \forall \config \in \confSet: \fSem \dimMeta = \t,
\elemSem \elem = \pElem}$.
As the formulation of grouping indicates, we can generate the group method for a 
variational entity once we have instantiated its configure method, i.e., grouping is the
extensional view of the configuration method. However, this would be
rather slow, hence, we define the group methods for individual instances of the variational 
interface in a more optimized manner. E.g., \figref{vq-group} defines grouping of v-queries 
and \secref{vra-sem} provides an example of it.



\subsection{SQL Generator}
\label{sec:sql-gen}

Independent from the approach used to translate a v-query to
a list of RA queries, 
to run the v-query on the underlying VDB
we need to generate SQL queries from RA queries. 
The SQL generator module achieves this by implementing 
two approaches:
1) generate an SQL query per RA query which is rather a straight 
forward translation
and 
2) generate an SQL query per a set of RA queries that requires 
generating a unified variational attribute set for all the RA queries, 
adjusting the projected attributes for each RA query to conform
to the unified variational attribute set by nullifying the attributes that 
do not exist in the original RA query,
and then unioning all the generated
SQL queries resulting in one SQL query.
SQL queries are generated w.r.t. SQL engine's type system,
e.g., a generated query has renaming for derived subqueries even 
if the RA query does not. 
%
For performance purposes, we use common table 
expressions (CTEs) to store temporary result and reuse them in
generated SQL queries.

\section{Experiments and Discussion}
\label{sec:exp-disc}

In this section, we discuss two real world use cases of variational databases:
1) Enron email use case from SPL
and 
2) employee use case from schema evolution)
We explain how we generated their VDBs and
the set of v-queries we used for our experiments. 
We then state different approaches that VDBMS provides for 
evaluating a v-query and the result of our experiments in \secref{apps}.
We finally conclude this section by a comprehensive discussion 
on benefits and shortcomings of VDBMS compared to current practices.

\subsection{Generating VDB}
\label{sec:gen-vdb}

As mentioned in \secref{impl}, VDBs are the input of VDBMS and 
since there are no tools to generate them, 
we manually generated the \emph{Enron email} and \emph{employee} 
VDBs from their two real world database variants. 

\textbf{Enron email VDB:}

\textbf{Employee VDB:}

\subsection{Experimental Queries}
\label{sec:exp-qs}

For performance purposes, we define two factors:
1) \emph{number of variants}, the number of relational queries
created from configuring a v-query
and
2) \emph{number of variations}, the number of relational queries
created from grouping a v-query.
To evaluate VDBMS using our two use cases, we consider a
comprehensive set of queries for each VDB. 

\textbf{Enron email query set:}

\textbf{Employee query set:}

\subsection{Approaches and Experiments}
\label{sec:apps}

There are four different SQL queries generated from an input set of RA 
queries by combining the approaches and optimizations. We use
different generators to simulate current approaches used to manage
variability within a context as well as comparing these methods with 
each other. The generated SQL queries need to be independent from the 
underlying DBMS that stores the VDB. Hence, the SQL generator module
has a submodule that prints generated SQL queries for each DBMS engine. 

\subsection{Discussion}
\label{sec:disc}

\section{Related Work}
\label{sec:rel-work}

We now survey works that address some kinds of variability in 
databases.

\textbf{Variational Research:}\\

\textbf{Schema Evolution:}\\

\textbf{Database Versioning:}\\


\textbf{SPL and SPL evolution:}\\


\section{Conclusion and Future Work}
\label{sec:conclusion}

%
%
%
%

\bibliographystyle{ACM-Reference-Format}
\bibliography{submission}
\appendix
\section{Feature Expression}
\label{app:fexp}
 \begin{figure}[H]
\textbf{Semantics of feature expressions:}
\begin{alignat*}{1}
\fSem [] . &: \ffSet \to \confSet \to \bSet\\
\fSem \bTag &= \bTag\\
\fSem \fName &= \config \ \fName\\
\fSem {\neg \fName} &= \neg \fSem \fName\\
\fSem {\annd \dimMeta} &= \fSem {\dimMeta_1} \wedge \fSem {\dimMeta_2}\\
\fSem {\orr \dimMeta} &= \fSem {\dimMeta_1} \vee \fSem {\dimMeta_2}\\
\end{alignat*}

\medskip
\textbf{Functions over feature expressions:}
\begin{alignat*}{1}
\mathit{sat}, \mathit{taut} &: \ffSet \to \bSet \\
\sat \dimMeta = \t &\textit{ iff \ } \exists \config \in \confSet: \fSem \dimMeta = \t\\
\taut \dimMeta = \t &\textit{ iff \ } \forall \config \in \confSet: \fSem \dimMeta = \t
\end{alignat*}

\caption{Feature expression evaluation and functions.}
\label{fig:fexp-eval}
\end{figure}

\section{V-Condition Configuration}
\label{app:vcond-form}
\begin{figure}[H]
\begin{alignat*}{1}
\ecSem [] . &: \vCondSet \to \confSet \to \pCondSet\\
\ecSem \bTag &= \bTag \\
\ecSem \vAttOpCte &= 
    \vAttOpCte\\
%
\ecSem \vAttOpAtt &= 
       \vAttOpAtt\\
%
\ecSem {\neg \vCond} &= \neg \ecSem \vCond\\
\ecSem {\orr \vCond} &= \ecSem {\vCond_1} \vee \ecSem {\vCond_2}\\
\ecSem {\annd \vCond} &= \ecSem {\vCond_1} \wedge \ecSem {\vCond_2}\\
\ecSem {\chc {\vCond_1, \vCond_2}} &=
	\begin{cases}
		\ecSem {\vCond_1}, &\text{ if } \fSem \dimMeta = \t \\
		\ecSem {\vCond_2}, &\text{ otherwise}
	\end{cases}
\end{alignat*}
\caption{Configuration semantics of v-conditions which assumes v-conditions
are well-typed.}
\label{fig:vcond-conf-sem}
\end{figure}

\section{Configure VDB}
\label{app:conf-vdb}

To acquire a specific variant of a VDB instance, as illustrated in 
the bottom of \figref{arch} and also known as 
VDB deployment in the context of SPL~\cite{vdbSpl18ATW}, 
we define the configuration semantics of VDB instances. To 
configure a VDB instance, we configure its v-schema and v-tables,
introduced in \appref{vsch-conf-sem} and \appref{vtab-conf-sem},
respectively.

\subsection{Configure V-Schema}
\label{app:vsch-conf-sem}

To configure a VDB, its v-schema needs to be configured as well.
V-schema configuration is a function that takes a v-schema and
a configuration and it returns the plain relational schema variant 
corresponding to the configuration. 
\figref{vsch-conf-sem} defines the configuration semantics
for v-schema and its elements, including variational attribute list
and relation schemas. 

\begin{figure}[H]

\textbf{Variational Attribute Set Configuration:}
\begin{flalign*}
& \olSem [] . : \vAttSet \to \confSet \to \pAttSet&
\end{flalign*}
\begin{flalign*}
& \olSem {\optAtt} &\spcEq&
    \begin{cases}
        \pAtt, \If \fSem {\dimMeta \wedge \getPC{\getRel \vAtt}} = \t\\
        \empAtt, \Otherwise
     \end{cases}\\
& \olSem {\optAtt, \vAttList} &\spcEq& \olSem {\optAtt}, \olSem \vAttList&\\
& \olSem \empAtt &\spcEq& \empAtt &
\end{flalign*}

\medskip
\textbf{V-Relation Configuration:}
\begin{flalign*}
&\orSem [] . : \vRelSchSet \to \confSet \to \pRelSchSet&
\end{flalign*}
\begin{flalign*}
&\orSem \vRelDef = 
	\begin{cases}
		\pRel \paran {\olSem {\vAttList}}, \If \fSem \dimMeta = \t\\
		\empRel, \Otherwise
	\end{cases}&
\end{flalign*}

\medskip
\textbf{V-Schema Configuration:}
\begin{flalign*}
&\osSem [] . : \vSchSet \to \confSet \to \pSchSet&
\end{flalign*}
\begin{flalign*}
&\osSem {\annot [\fModel] {\setDef {\vRelDefNum 1, \ldots, \vRelDefNum n}}}\\
&\hspace{0.3cm}= \begin{cases}
                 \setDef {\orSem {\vRel_1\left( \vAttList_1 \right)^{\dimMeta_1 \wedge \fModel} }, \ldots, 
                 \orSem {\vRel_n\left( \vAttList_n \right)^{\dimMeta_n \wedge \fModel} }},		
        \text{ if } \fSem \fModel = \t\\
        \setDef \ ,\qquad\qquad\qquad\qquad\qquad\quad \text{ otherwise}
	\end{cases}&\\
\end{flalign*}

\caption{
Configuration of variational set of attributes, v-relations, and v-schema.
\empAtt\ denotes both an empty variational set of attributes and a non-existence 
relation.
Note that the feature model and 
relation presence condition are passed all the way to attributes due to the 
hierarchal structure of presence conditions within a v-schema.}
\label{fig:vsch-conf-sem}
\end{figure} 

\subsection{Configure V-Table}
\label{app:vtab-conf-sem}

To configure a VDB, its content, i.e., tables, need to be configured too.
The relation schemas of tables have been configured by the 
configuration of v-schema, hence, the configuration of v-table
needs to only configure tuples. I.e., v-table configuration already
knows the schema of the relation (i.e., its attributes), hence, it
does not need to check if a value should be included or not. 
\figref{vtab-conf-sem} defines v-table configuration semantics. 

\begin{figure}[H]
\textbf{V-Tuple Configuration:}
\begin{flalign*}
&\ouSem [] . : \vRelCont \to \confSet \to \pRelCont&
\end{flalign*}
\begin{flalign*}
&\ouSem {\left( {\vi v k, \dimMeta_\tuple}\right)} = \left( \ovSem {v_1}, \hdots, \ovSem {v_k} \right) &\\
& \textit{ where } \forall 1 \leq i \leq k: \ovSem {v_i} = 
\begin{cases}
v_i, \If \fSem {\getPC {\getAtt {v_i}} \wedge \dimMeta_\tuple} = \t\\
\varepsilon, \Otherwise
\end{cases}
\end{flalign*}

\medskip
\textbf{V-Table Configuration:}
\begin{flalign*}
&\otSem [] . : \vRelContSet \to \confSet \to \pRelContSet&
\end{flalign*}
\begin{flalign*}
&\otSem {\setDef {\vi \tuple n}} = \setDef {\ouSem {\tuple_1}, \hdots, \ouSem {\tuple_n}}&
\end{flalign*}

\medskip
\textbf{VDB Instance Configuration:}
\begin{flalign*}
&\odbSem [] . : \vInstSet \to \confSet \to \pInstSet&
\end{flalign*}
\begin{flalign*}
&\odbSem {\annot [\fModel] {\setDef {\vi \vRelCont l}}} = \setDef {\otSem {\vRelCont_1}, \hdots, \otSem {\vRelCont_l} }&
\end{flalign*}

\caption{V-tuple, v-table, and VDB instance configuration.
$\varepsilon$ denotes a non-existence value. }
\label{fig:vtab-conf-sem}
\end{figure}

\end{document}